\pdfoutput=1

\documentclass[11pt]{article}
\usepackage{multirow}
\usepackage{fancyvrb}
\usepackage{fvextra}
\usepackage{times}
\usepackage{ACL2023}
\usepackage{latexsym}
\usepackage{textcomp}
\usepackage{booktabs}
\usepackage{svg}
\usepackage{graphicx}
\usepackage{amsmath}
\usepackage{amssymb}
\usepackage{amsfonts}
\usepackage{enumitem}
\usepackage{xcolor}

\DefineVerbatimEnvironment{CustomVerbatim}{Verbatim}{fontsize=\small, frame=single, breaklines=true, breaksymbolleft={}}

\usepackage[T1]{fontenc}

\usepackage[utf8]{inputenc}

\usepackage{microtype}
\usepackage{tikz}
\newcommand*\circled[1]{\tikz[baseline=(char.base)]{
            \node[shape=circle,draw,inner sep=2pt] (char) {#1};}}

\usepackage{inconsolata}

\title{CRScore: Grounding Automated Evaluation of Code Review Comments in Code Claims and Smells}

\author{Atharva Naik ~~ Marcus Alenius ~~ Daniel Fried ~~ Carolyn Ros\'e \\
  Language Technologies Institute \\
  Carnegie Mellon University \\
  \texttt{\{arnaik, malenius, dfried, cprose\}@cs.cmu.edu}}

\begin{document}
\maketitle
\begin{abstract}
The task of automated code review has recently gained a lot of attention from the machine learning community. 
However, current review comment evaluation metrics rely on comparisons with a human-written reference for a given code change (also called a \textit{diff}).  Furthermore, code review is a one-to-many problem, like generation and summarization, with many ``valid reviews'' for a diff.
Thus, we develop \texttt{CRScore} --- a reference-free metric to measure dimensions of review quality like conciseness, comprehensiveness, and relevance.
We design \texttt{CRScore} to evaluate reviews in a way that is grounded in claims and potential issues detected in the code by LLMs and static analyzers.
We demonstrate that CRScore can produce valid, fine-grained scores of review quality that have the greatest alignment with human judgment among open source metrics (0.54 Spearman correlation) and are more sensitive than reference-based metrics.
We also release a corpus of 2.9k human-annotated review quality scores for machine-generated and GitHub review comments to support the development of automated metrics\footnote{\url{https://github.com/atharva-naik/CRScore}}.


\end{abstract}

\section{Introduction}
Code Review is an essential quality control tool for software engineers to ensure that source code is free of bugs and upholds standards \cite{codereview_impact, four_eyes}.
Software engineers prefer lightweight, asynchronous review processes, as enabled through GitHub's review comment feature, over formal, in-person reviews \cite{ModernCodeReview, Badampudi_2023}.
This has led to the creation of benchmarks for automated generation of natural language (NL) review comments \cite{CodeReviewer, tufano2022using}.
However, these benchmarks use reference-based evaluation metrics like BLEU \cite{BLEU}, which have been shown to have low validity \cite{reiter-2018-structured, out_of_the_bleu}, especially when paired with limited and low-quality references. 

\begin{figure}[!tbh]
    \centering
    \includegraphics[width=0.5\textwidth]{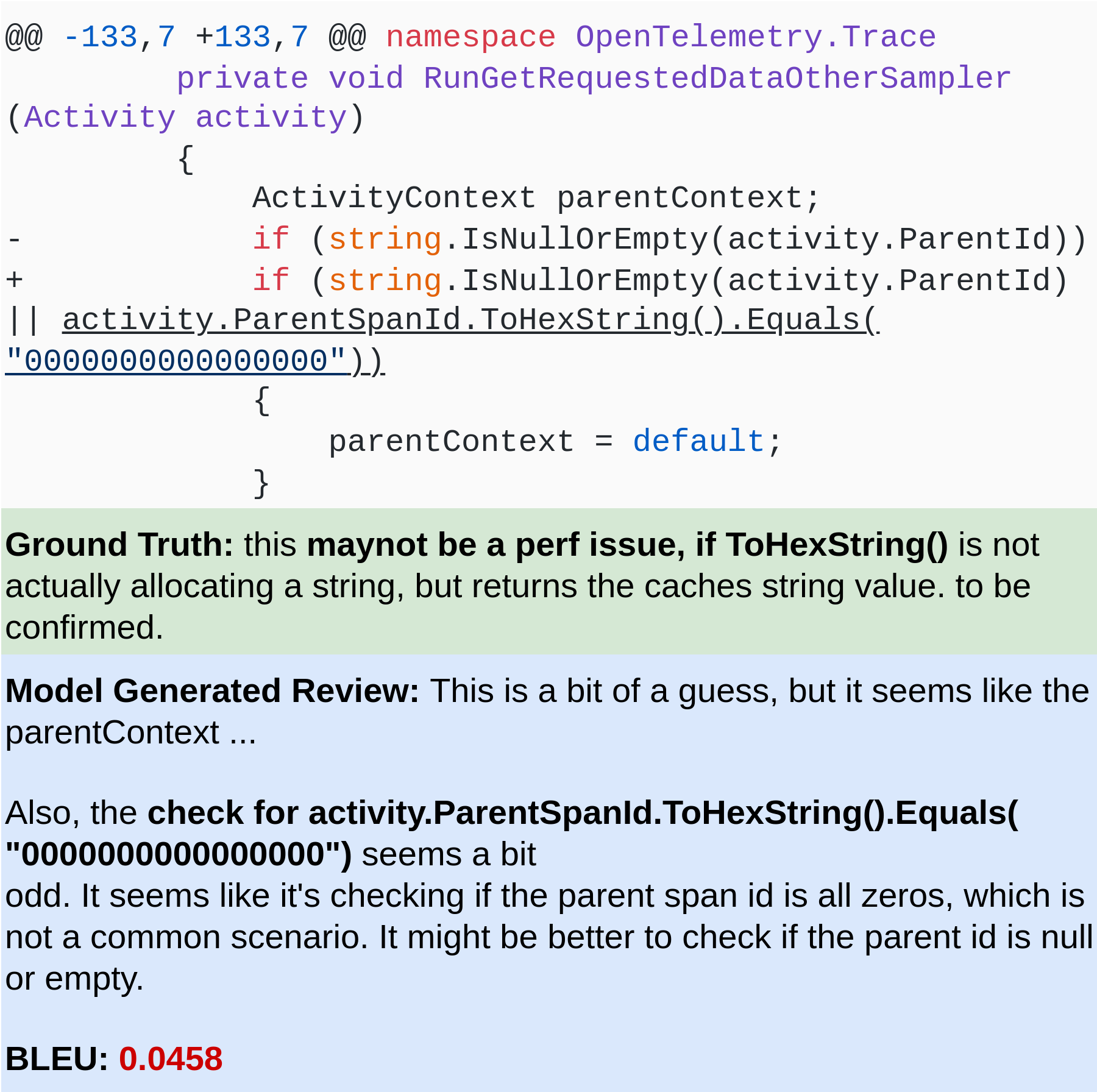}
    \caption{Example diff with multiple valid reviews. The ground truth and model-generated reviews focus on different topics, like the performance of the added check, and how likely it is to be triggered. However, a reference-based metric like the BLEU score assigns this review a low score of 0.0458.}
    \label{fig:multiple_valid_reviews}
\end{figure}

Code review is fundamentally a one-to-many problem, where a given diff can have multiple possible issues that a review can tackle. 
Having a limited number of reference reviews (e.g., one per diff in \texttt{CodeReviewer} \cite{CodeReviewer}) leads to unfairly low scores with reference-based metrics.
For example, for the diff shown in Figure~\ref{fig:multiple_valid_reviews}, the ground truth review focuses on whether the \texttt{ToHexString()} function could cause a performance issue. However, the model-generated review focuses on \texttt{ToHexString().Equals("0000000000000000")} being an odd condition with a scenario where its being triggered is unlikely, which is also a valid review for the diff.
However, the BLEU score value for the model-generated review is very low, specifically at 0.0458, due to poor n-gram overlap. 
Additionally, the references can also be low-quality due to missing context, as shown in Table~\ref{tab:missing_context_eg}, or can focus on trivial and tangential issues \cite{review_comments_discuss_tangential_issues}.
Such low-quality references, paired with reference-based metrics, can harshly and unfairly penalize models. 

Motivated by these drawbacks, we propose \texttt{CRScore}, an automated but reference-free evaluation metric that uses dimensions of review quality from prior work \cite{piorkowski2020towards, turzo2024makes}. 
In particular, \textit{Comprehensiveness} -- does the review convey all the necessary information? \textit{Conciseness} -- does the review only convey the necessary information in an efficient way and \textit{Relevance} -- is all the information on topic? 
We operationalize our metric through a two-step process: \circled{1} generate a list of \emph{pseudo-references} spanning information like possible claims, issues, and implications of a code change, and \circled{2} use semantic textual similarity (STS) to align parts of the review to the pseudo-references.
To generate the pseudo-references, we use a neuro-symbolic approach that combines \textbf{Large Language Models} (LLMs) and \textbf{Code Analysis Tools} (CATs) that can detect formatting errors, faulty design patterns (code smells \citealt{rasheed2024ai}), etc.                                We combine these methods for more exhaustive pseudo-references and to overcome drawbacks of each method (section~\ref{sec:benefits_of_neurosymbolic_combination}).   


Finally, we demonstrate the validity of \texttt{CRScore} by human evaluation of the quality of pseudo-references and by measuring the alignment of our metrics with human judgment.
We show that a large number (82.6\%) of pseudo-references generated by \texttt{CRScore} are correct, and that it has the greatest alignment with (Spearman correlation 0.5431) and sensitivity to (Figure~\ref{fig:metric_monotonicity}) the human judgment of review quality, as compared to that of the reference-based metrics.
\textbf{Contributions.}
(1) We propose an automated reference-free metric that combines the advantages of LLMs and CATs to measure review quality along fine-grained dimensions.
(2) We collect human annotations of pseudo-reference quality and review conciseness, comprehensiveness, and relevance to validate our approach. We plan to make these scores publicly available as a resource for continued development of automated metrics.
(3) We benchmark several LLMs of code for code review on the \texttt{CodeReviewer} dataset using \texttt{CRScore} and compare it with reference based metrics.

\section{Related Work}

In this section, we summarize the limitations of reference-based evaluation, the need for better code review evaluation metrics, inspiration from reference-free evaluation for other tasks, and how code smells can be leveraged to evaluate reviews.

\subsection{Reference Based Evaluation Metrics:}
\label{sec:related_work:ref_based_eval}
Reference-based metrics like BLEU \cite{BLEU}, ROUGE \cite{ganesan2018rouge}, and BERTScore \cite{BERTScore} have seen widespread adoption for text generation tasks like translation and summarization due to their convenience.
While metrics like BLEU, ROUGE, and character F-score \cite{chrF, chrF++} use n-gram overlap between the reference and candidate text, metrics like BERTScore \cite{BERTScore} try to capture the semantic similarity. 
However prior studies have shown metrics like BLEU to have low validity (overlap with human judgment) and reliability \cite{reiter-2018-structured, out_of_the_bleu}.
Meanwhile, BERTScore can fail for candidates with errors that are lexically and stylistically similar to references \cite{hanna-bojar-2021-fine}. 

\subsection{Code Review Evaluation}
Due to the high time and resource demands of code review, automated approaches have gained popularity \cite{yang2024survey}. 
\citet{tufano2021towards, tufano2022using, CodeReviewer} proposed large datasets for code review tasks like code changes quality detection, review comment generation, and code refactoring.
However, these datasets used reference-based metrics and thus suffer from the issues highlighted in sec~\ref{sec:related_work:ref_based_eval}. 
While many studies have focused on modeling methods for code review tasks \cite{PORNPRASIT2024107523, LLaMAReviewer, dong2024gpt, fan2024exploring, finetune_code_review_comprehensibility, lin2024improving}, they either retain the same reference-based automated metrics like BLEU \cite{BLEU}, or use human evaluation.
Some studies have focused on evaluating style and presentation of review comments for usefulness \cite{RevHelper, CommentBERT, EvaCRC} (see  Appendix~\ref{sec:additional_related_work:review_presentation_analysis} for detailed comparison) or negativity and toxicity \cite{ahmed2017senticr, sarker2023automated}.
This work focuses on content-focused and reference-free automated evaluation. 
We show that reference-based metrics combined with noisy references fail to capture human preferences.
We propose \texttt{CRScore}, the first automated content-focused reference-free metric to overcome these limitations.

\subsection{Reference Free Evaluation}
Reference-free evaluation metrics have been proposed for various text-generation tasks to capture multiple valid outputs.
Instead of using references, these metrics try to measure general ``quality dimensions'' like relevance, informativeness, etc.
VIFIDEL \cite{VIFIDEL}, InfoMetIC \cite{hu2023infometic}, and ClipScore \cite{ClipScore} evaluate dimensions like faithfulness, informativeness, and relevance, respectively for image captioning. 
FED \cite{mehri2020unsupervised}, 
and USR \cite{mehri-eskenazi-2020-usr} evaluate dimensions like informativeness, relevance, and overall quality for dialog.
Studies on the helpfulness of software documentation and code review \cite{piorkowski2020towards, turzo2024makes} propose quality dimensions like conciseness, completeness, understandability, relevance, and supporting evidence.
The common trend across these studies is some notion of conciseness, informativeness/comprehensiveness, and relevance being useful, prompting us to focus on them. 
However, our work is the first to operationalize these dimensions for automated evaluation.

\subsection{Code Smell Detection}
``Code smells'' \cite{fowler1997refactoring} are design flaws and bad practices (also called \textit{anti-patterns}) that can lead to maintainability issues.
Detecting code smells automatically has traditionally been accomplished by static analysis-based approaches \cite{JDeodorant, Paiva2017, liu2017detection}. 
Recently machine learning \cite{PyCodeSmell} and transfer learning \cite{CodeSmellDLTL} based approaches have been proposed to learn more complex heuristics.
Recent approaches have leveraged LLMs via prompting \cite{liu2024prompt} and agents \cite{rasheed2024ai} to achieve improvement and tackle repository-level code smell detection.
However, these approaches are limited in the types of smells they target \cite{liu2024prompt, PyCodeSmell}, training data requirements \cite{zhang2024datapreparationdeeplearning}, or lack comprehensive evaluation \cite{rasheed2024ai}.
Also, code smells differ across programming languages \cite{code_smells_multi_ling}, and transfer learning approaches can only be leveraged for similar languages \cite{CodeSmellDLTL}.
Due to these limitations of learning-based methods and to mitigate the self-selection bias of LLMs (sec \ref{sec:benefits_of_neurosymbolic_combination}) we use code analysis tools.

\section{Operationalizing \texttt{CRScore}}
\begin{figure*}[!tbh]
    \centering
    \includegraphics[width=1\textwidth]{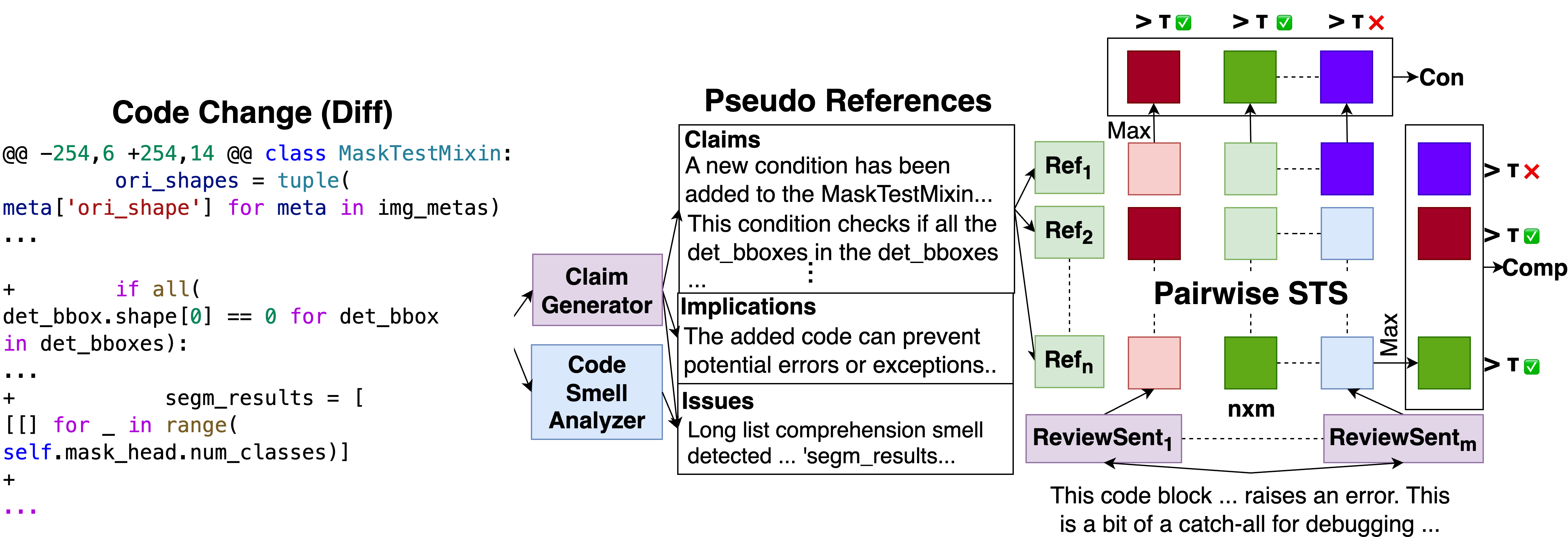}
    \caption{\textbf{Operationalization of \texttt{CRScore}:} Our metric first generates pseudo-references for the diff --- claims, implications and issues. Then each pseudo-reference is embedded by a sentence transformer along with each review sentence and the pairwise semantic textual similarity (STS) is computed. The high similarity threshold $\tau$ is used to compute the Con and Comp metrics whose harmonic mean yields the Rel score.}
    \label{fig:crscore_pipeline}
\end{figure*}

Motivated by the one-to-many nature of code reviews, noisy references, and the pitfalls of reference-based automated metrics, we develop CRScore --- a reference-free, quality dimension-based automated metric.            
As shown in Figure~\ref{fig:crscore_pipeline}, instead of relying on explicit references, our metric generates ``pseudo-references'' from the code change spanning claims, implications, and issues/smells that hurt maintainability --- in other words, some topics that a review should address \cite{rasheed2024ai}.  
Then we use semantic textual similarity (STS) measures to quantify how much these topics are addressed by a code review as shown in Figure~\ref{fig:crscore_pipeline} through the lens of three quality dimensions: conciseness, comprehensiveness, and relevance. 
They capture review quality, similar to precision, recall, and f-score used for classification and retrieval.
We describe the three main components of our framework --- the quality dimensions, pseudo-reference generation, and similarity measurement --- below.

\subsection{Quality Dimensions}
\label{sec:eval_framework}
We are inspired by \cite{piorkowski2020towards, turzo2024makes} to pick the dimensions of ``conciseness'', ``comprehensiveness'' (similar to completeness), and ``relevance''.
The former two capture and strike a balance between comprehensive reviews with a lot of detail and concise, minimalist reviews.
Conciseness and comprehensiveness play the role of precision and recall respectively, capturing how much of the review is on topic, and how much information is covered.
Relevance strikes a balance between the two, like f-score, measuring the overall quality of the review.
Additionally, we also do a human evaluation of the validity of the pseudo-references focusing on aspects like supporting evidence \cite{piorkowski2020towards}.

\subsection{Pseudo Reference Generation}
To generate pseudo references we develop an LLM-based pipeline for generating \textbf{claims} about the code changes on two levels of abstraction: 1) Low-level changes and 2) High-level ``implications'' of the change.
To allow for reproducibility and ease of deployment we use a 6.7B parameter open source LLM, Magicoder \cite{wei2023magicoder}. 
We fine-tune Magicoder-S-DS-6.7B for this task using task-specific data produced by a stronger model (GPT-4) as shown in Figure~\ref{fig:claim_gen_pipeline}.
We generate claims by prompting GPT-4 for a random subset of 1k code changes from the CodeReviewer validation data.

In addition to the claims, we also utilize program analysis tools that can detect \textbf{issues} and ``code smells'' (characteristics that indicate deeper flaws in programs as shown in Table~\ref{tab:pyscent}) like PySmell\footnote{\url{https://github.com/whyjay17/Pyscent}} as well as static analyzers like PMD\footnote{\url{https://pmd.github.io/}} and JSHint\footnote{\url{https://jshint.com/}}. 
They can detect unused variables, unnecessary object creation, syntax errors, leaking variables, type conversion issues, etc. 
These tools target Python, Java, and Javascript respectively, and can use rule-based analysis to detect issues at a file or repository level.
They complement the aspects that might be missed by an LLM-based analysis.
We combine the code smells and claims to get the final set of pseudo references.

\subsection{Computing Similarity with Pseudo References:}
We use a Sentence Transformer \cite{reimers2019sentence} model (\texttt{mxbai-embed-large-v1}\footnote{\url{https://huggingface.co/mixedbread-ai/mxbai-embed-large-v1}}) to compute Semantic Textual Similarity (STS) between pseudo-references and review sentences. 
The pairwise similarities are then used to compute the conciseness, comprehensiveness, and relevance, as shown by equations~\ref{eq:con}, \ref{eq:comp}, and \ref{eq:rel}.
We picked this model because it has the best performance, as of July 2024, on English STS on the MTEB benchmark \cite{muennighoff2022mteb} for models with less than 1B parameters.

We start by computing token embeddings for the pseudo-references (p-refs) and review sentences (r-sents), excluding the stopwords and pooling the rest of the token embeddings to build representations for the whole sentence.
Then we compute STS scores with these sentence embeddings via pair-wise cosine similarity $s()$ between the p-refs ($\mathcal{P}$) and r-sents ($\mathcal{R}$).
The Conciseness ($Con$) which is computed as:
\begin{equation}
    Con = \frac{\sum_{r \in \mathcal{R}}I[\max_{p \in \mathcal{P}}s(c,r)>\tau]}{|\mathcal{R}|}
\label{eq:con}
\end{equation}
represents the fraction of r-sents from the model-generated review with greater similarity to any p-ref above a threshold $\tau$.
Here, $I$ is an indicator variable such that:
\[
I[x]=\begin{cases}
			1, & \text{if $x$ is true}\\
            0, & \text{otherwise}
		 \end{cases}
\]
$Con$ resembles precision as it captures the fraction of r-sents (candidate set) that are ``on topic'' concerning the p-refs (reference/gold set).
The Comprehensiveness ($Comp$) computed as:
\begin{equation}
    Comp = \frac{\sum_{p \in \mathcal{P}}I[\max_{r \in \mathcal{R}}s(c,r)>\tau]}{|\mathcal{P}|}
\label{eq:comp}
\end{equation}
represents the fraction of p-refs that have greater similarity to any of the r-sents than the threshold $\tau$.
This metric resembles recall as it captures the fraction of the p-refs (gold set) covered by the model-generated r-sents (candidate set).
There is a trade-off between conciseness and comprehensiveness just like precision and recall, and to capture the trade-off between these metrics like the F1 score, we define the Overall Relevance $Rel$ as the harmonic mean of $Con$ and $Comp$:
\begin{equation}
    Rel = \frac{2 \cdot Con\cdot Comp}{Con+Comp}
\label{eq:rel}
\end{equation}
Based on the definition of $I[x]$ all of our metrics range in value from $[0,1]$.
\\
\textbf{Selecting threshold $\tau$ for high similarity:}
\label{sec:selecting_threshold_for_similarity}
For our similarity threshold $\tau$ we use the average similarity between a pseudo-reference and the review sentence talking about it.
To compute it we use the distribution of pseudo-reference and review similarity scores for the best model (GPT-3.5 in our case).
Starting with the \texttt{CodeReviewer} test set we exclude the data used for collecting human annotations. 
For the remaining 9869 instances, we compute the similarity between each review sentence and pseudo-reference for its corresponding code change.
Then we associate each review sentence with the most similar pseudo-reference to capture correspondence.
Finally, we average the similarity score value across all review sentences, giving us a value for the threshold\footnote{We also show that computing the threshold with ground truth reviews yields a comparably good threshold $\tau_{GT} = 0.6576$ with respect to correlation with human annotations.} $\tau_{best} = 0.7314$. 
While choosing the right threshold is important for our metric, we note that it is robust to some variations in the value as shown in Table~\ref{tab:STS_threshold_variation}.

\section{Validating \texttt{CRScore}}
To be a valid metric, \texttt{CRScore} needs to satisfy a few properties.
Firstly, the generated pseudo-references should have few errors and unverifiable claims while being exhaustive. 
Additionally, we want our dimension-level scores ($Con$, $Comp$, and $Rel$), especially $Rel$, to correlate with human judgment for each dimension. 
The most important property, is arguably the ability of our metric (especially $Rel$) to rank review generation systems in the same order as humans.  
In the subsequent sections, we describe how we design the experiments to collect annotations for these properties (section~\ref{sec:phase1}, \ref{sec:phase2}), choose a set of systems to be rated by humans (section~\ref{sec:exp_systems}), and set up reference-based metrics (section~\ref{sec:exp_metrics}) for comparison \texttt{CRScore}.

\subsection{Rating Quality of Pseudo-References}
\label{sec:phase1}
To show that LLM-generated pseudo-references used by the \texttt{CRScore} evaluation pipeline are high-quality, we gather annotations capturing incorrect, unverifiable, and missing claims.
We had two trained human annotators (co-authors of this paper) judge the quality of the pseudo-references for 100 randomly sampled code changes from \texttt{CodeReviewer} \cite{CodeReviewer} each in Python, Java, and Javascript (300 total, see Appendix~\ref{sec:dataset_stats} for details).
The annotators were asked to code the pseudo-references as 1 (correct based on evidence), 0 (incorrect based on evidence), and -1 (unverifiable due to lack of evidence).
They were also asked to add any pseudo-references about issues/claims not covered by the pseudo-references.
We report the fraction of correct claims (accuracy), incorrect claims (error rate), unverifiable claims (unverifiable rate), and missing claims (missing rate).
If $N_c$, $N_u$, $N_i$, and $N_a$ represent the number of correct, unverifiable, incorrect, and added claims then each of these rates can be calculated as:
\[
\text{Accuracy} = \frac{N_c}{N_c+N_u+N_i}
\]
\[
\text{Error Rate} = \frac{N_i}{N_c+N_u+N_i}
\]
\[
\text{Unverifiable Rate} = \frac{N_u}{N_c+N_u+N_i}
\]
\[
\text{Missing Rate} = \frac{N_a}{N_c+N_u+N_i}
\]
Based on these expressions: $\text{Accuracy}+\text{Unverifiable Rate}+\text{Error Rate}=1$.
The results for each language are shown in table~\ref{tab:human_study_phase1}.
The annotators follow guidelines laid out in a codebook (section~\ref{sec:codebook_for_rating_pseudo_ref_quality}) which includes examples for each category. 
We measured the coding reliability of our approach by collecting annotations from both annotators on a common set of 100 pseudo-references. 
These annotations yielded a Cohen Kappa of 0.804 which indicates great inter-annotator reliability \cite{Landis1977-qw}.
Also, we only evaluate the LLM-generated claims here as we know the static code analysis tools are rule-based and reliably correct. 

\subsection{Rating Review Quality Dimensions}
\label{sec:phase2}
To show that \texttt{CRScore} aligns with the human judgment of review quality along the proposed dimensions: comprehensiveness, conciseness, and relevance we gather annotations from the same annotators on reviews generated by 9 systems (section~\ref{sec:exp_systems}) and the ground truth references. 
We use the same code changes used in section~\ref{sec:phase1}.

The raters again follow a codebook (section~\ref{sec:rating_review_quality}) that contains guidelines and examples for annotating each dimension using the pseudo-references on a 5-point Likert scale.
The raters use the updated set of pseudo-references based on the first phase of annotations described in section~\ref{sec:phase1}. 
Incorrect and unverifiable references are removed and missing claims are added.
We also add the pseudo-references (issues and code smells) generated by the code analysis tools for each code change.
Raters are also asked to annotate any claims they find unnecessary for a given code change, which are then excluded while rating the review on each quality dimension.
Again, we measure the reliability of the codebook by collecting annotations from the two raters, this time on a common set of 100 reviews. 
We compute Krippendorff's alpha reliability \cite{krippendorff2018content} for each dimension, yielding values of 0.8868, 0.8505, and 0.8806 for conciseness, comprehensiveness, and relevance respectively. 
Alpha values > 0.8 are generally considered reliable.
Despite the steps taken to ensure reliability of the annotations, their might be some concerns about potential biases, which we address in Appendix~\ref{sec:appendix:bias_in_phase2}.

Based on these annotations, we measure the agreement between \texttt{CRScore} ($Rel$) and the reference-based metrics with the human-annotated relevance Likert scores using Kendal and Spearman correlation (Table~\ref{tab:metric_correlations}). 
We also compute the correlation between the system rankings generated by the metric and human annotations (Table~\ref{tab:system_ranking_correlations}).

\begin{table}[!tbh]
\centering
\resizebox{0.5\textwidth}{!}{
\begin{tabular}{@{}lrrrr@{}}
\toprule
\textbf{Language} &
  \textbf{Accuracy} &
  \textbf{\begin{tabular}[c]{@{}r@{}}Error\\ Rate\end{tabular}} &
  \textbf{\begin{tabular}[c]{@{}r@{}}Unverifiable\\ Rate\end{tabular}} &
  \textbf{\begin{tabular}[c]{@{}r@{}}Missing\\ Rate\end{tabular}} \\ \midrule
Python     & 83.65 & 12.02 & 4.33  & 5.77 \\
Java       & 79.09 & 13.22  & 7.69  & 8.89 \\
Javascript & 85.21 &  6.52  & 8.27 & 2.51 \\ \bottomrule
\end{tabular}
}
\caption{Quality of pseudo-references -- the fraction of correct (accuracy), incorrect (error rate), unverifiable (unverifiable rate), and added claims (missing rate) based on human annotations.}
\label{tab:human_study_phase1}
\end{table}

\begin{table}[!tbh]
\centering
\resizebox{0.48\textwidth}{!}{
\begin{tabular}{@{}lrrrr@{}}
\toprule
\textbf{Metric} &
  \multicolumn{1}{r}{\textbf{$r_s$}} &
  \multicolumn{1}{r}{$p$} &
  \multicolumn{1}{r}{\textbf{$\tau$}} &
  \multicolumn{1}{r}{$p$} \\ \midrule
BLEU                                                               & -0.3          & 0.433            & -0.1667         & 0.612            \\
\begin{tabular}[c]{@{}l@{}}BLEU\\(without stop)\end{tabular}                                               & -0.15         & 0.7              & -0.0556         & 0.919            \\
BERTScore                                                          & 0.35          & 0.356            & 0.2222          & 0.477            \\
\begin{tabular}[c]{@{}l@{}}Normalized\\ Edit Distance\end{tabular} & 0.1667        & 0.668            & 0.0556          & 0.919            \\
\begin{tabular}[c]{@{}l@{}}ROUGE-L\\ F-measure\end{tabular}        & 0.0167        & 0.966            & 0.556           & 0.919            \\
chrF                                                               & 0.4833        & 0.187            & 0.3889          & 0.18             \\
chrF++                                                             & 0.6           & 0.088            & 0.4444          & 0.119            \\
LaaJ-Magic                                                             & 0.8           & 0.010            & 0.6667          & 0.013            \\
\textbf{LaaJ-GPT}                                                            & \textbf{0.9833}          & \textbf{1.9e-6}           & \textbf{0.9444}          & \textbf{5.0e-5}            \\
\begin{tabular}[c]{@{}l@{}}\textbf{Rel} ($\tau_{best}$)\\\textbf{ours}\end{tabular} & \underline{0.95} & \underline{8.7e-5} & \underline{0.8889} & \underline{2.4e-4} \\ \bottomrule
\end{tabular}
}
\caption{
Spearman ($r_s$) and Kendall ($\tau$) correlations between system rankings produced by each metric and human annotations for relevance. 
Only our metric and the LaaJ variants achieve a high, statistically significant correlation. Our metric has a comparable correlation to LaaJ-GPT which uses a much more powerful closed-source model, while being much better than LaaJ-Magic which uses Magicoder-S-DS-6.7B, the same based model as CRScore.
}
\label{tab:system_ranking_correlations}
\end{table}

\subsection{Annotating Similarity between Pseudo-References and Reviews}
While annotating the review quality dimensions using the pseudo-references the annotators are also asked to mention the pseudo-references covered by each system-generated review. 
This gives us a dataset of 5.7k pseudo-reference and review pairs, where 1948 are positive examples or cases where the review addresses a claim according to a human, while the rest are negative examples or pairs where the reviews don't address a pseudo-reference. 
Using this data we evaluate the STS model with various similarity thresholds ($tau$) using it to covert the similarity score into a binary classifier and report its precision, recall and F1-score in predicting whether in a pseudo-reference and review pair the review addresses the pseudo-reference.
The results are shown in Table~\ref{tab:STS_eval}

\subsection{Review Generation Systems}
\label{sec:exp_systems}
To see if \texttt{CRScore} can rank code review systems of varying capacity, we choose a diverse set of review generation models. 
They span various parameter sizes, pre-training, and fine-tuning strategies: \\
\textbf{Simple baselines:} We create two simple baselines, namely, a BM-25 retriever and an LSTM as described in section~\ref{sec:simple_baselines}. 
We choose these models with the expectation that they will likely perform the worst, to see if our metric assigns them a low score. \\
\textbf{CodeReviewer:} We pick the CodeReviewer model from \cite{CodeReviewer} as it is a transformer-based model trained on code review-specific data and objectives. \\
\textbf{Open source LLMs}: We prompt several open-source LLMs in a few-shot manner with a fixed set of three example code changes and review pairs from the validation set. 
We use LLMs in the 3-13B parameter range: Stable-Code-3B \cite{stable-code-3b}, DeepSeekCoder-6.7B \cite{guo2024deepseek}, Magicoder-6.7B \cite{wei2023magicoder}, CodeLLaMA-7B and 13B \cite{roziere2023code} and LLaMA-3-8B \cite{llama3modelcard}. \\
\textbf{Closed source LLMs:} We prompt closed-source LLMs like GPT-3.5 in a manner similar to the open-source LLMs.


\subsection{Reference-based Metrics}
\label{sec:exp_metrics}
We pick commonly used reference-based metrics for code review and other text generation tasks to compare with \texttt{CRScore}: \\
\textbf{BLEU} \cite{BLEU} measures the n-gram precision between the generated text and references with an additional brevity penalty to discourage short outputs.
It is used for evaluation in both \cite{tufano2021towards} and \texttt{CodeReviewer} \cite{CodeReviewer}.
We report the results with and without stop word removal. \\
\textbf{Normalized Edit Distance} is a normalized Levenshtein distance used in prior work \cite{tufano2021towards, bairi2024codeplan} to measure the number of edits required to match candidate and target reviews or code. \\
\textbf{ROUGE-L F-measure} is a popular recall-oriented metric originally proposed for summarization and machine translation.
We use the longest common subsequence-based sentence level f-measure implementation.\footnote{\url{https://pypi.org/project/rouge-score/}} \\
\textbf{chrF:} \cite{chrF} Is a machine translation metric which is essentially a character level F score computed using character level n-grams. \\
\textbf{chrF++:} \cite{chrF++} Is a variant of chrF that additionally incorporates word n-grams. \\
\textbf{BERTScore:} \cite{BERTScore} We use the BERTScore F1 measure to capture the semantic similarity between the reference review and the generated review. \\
\textbf{LLM-as-a-judge (LaaJ):} Following recent developments \cite{zheng2023judging} we develop an LLM prompting-based approach that compares the model-generated reviews against \texttt{CodeReviewer} references to generate relevance scores. We prompt them with descriptions of the same Con, Comp, and Rel dimensions developed in section~\ref{sec:eval_framework} with the prompt shown in~\ref{sec:appendix:llm_as_a_judge_prompt}. We evaluate two variants of this metric one that uses GPT-4o as the judge LLM (LaaJ-GPT) and one that uses Magicoder-S-DS-6.7B (LaaJ-Magic), the opensource model used by CRScore, for a fair comparison.

\section{Results}
\subsection{Validity of Pseudo-References}
We show the rates of correct, incorrect, unverifiable, and missing claims as explained in section \ref{sec:phase1} in Table~\ref{tab:human_study_phase1}.
The pseudo-references produced by our pipeline were relatively accurate with roughly 82.6\% accuracy across the languages.
The best performance is for Javascript and the worst is for Java.
The most frequent issues in the code claims were incorrect claims. 
Most of the errors were in code comprehension (``misreading the code'') and over-generalization (incorrect assumptions made from the limited context, contradicting file level context). 
Some examples are shown in Table~\ref{tab:pseudo_reference_errors}.
For Javascript, we observed unverifiable claims to be the biggest issue, e.g., claims about code efficiency or functionality made without evidence. 
E.g.: ``However, this change could also potentially enable less strict \dots \textbf{behavior}, \dots This could make the code \textbf{less efficient} \dots''.


\begin{table}[!tbh]
\centering
\begin{tabular}{@{}lrr@{}}
\toprule
\textbf{Metric}                                                    & \multicolumn{1}{l}{$\tau$} & \multicolumn{1}{l}{$r_s$} \\ \midrule
BLEU                                                               & \textcolor{gray}{0.001}               & \textcolor{gray}{-0.0001}               \\
\begin{tabular}[c]{@{}l@{}}BLEU\\ (without stopwords)\end{tabular} & 0.0425   & 0.0542               \\
BERTScore                                                          & 0.081 & 0.1083             \\
\begin{tabular}[c]{@{}l@{}}Normalized\\ Edit Distance\end{tabular} & \textcolor{gray}{0.0193} & \textcolor{gray}{0.0249}               \\
\begin{tabular}[c]{@{}l@{}}ROUGE-L\\ F-measure\end{tabular} & 0.0757 & 0.0989               \\
chrF & 0.1484 & 0.1966 \\
chrF++ & 0.1555 & 0.2057 \\
LaaJ-Magic & 0.2464 & 0.2748 \\
\textbf{LaaJ-GPT} & \textbf{0.5247} & \textbf{0.605} \\
\textbf{Rel ($\tau_{best}$) (Ours)} & \underline{0.4567} & \underline{0.5431}    \\ \bottomrule
\end{tabular}
\caption{Comparing Kendal-Tau ($\tau$) and Spearman Rank ($r_s$) correlation of reference-based evaluation metrics and our reference-free relevance score (Rel) with human annotations for the relevance dimension. Results that don't achieve statistical significance are grayed out. 
Our metric archives the second-best correlation behind LaaJ-GPT however it does so with a much smaller 6.7B parameter open-source model. 
Additionally, CRScore outperforms LaaJ-Magic which uses the same base LLM (Magicoder) as the judge model.
}
\label{tab:metric_correlations}
\end{table}

\subsection{Validity of Review Quality Dimension Scores (Con, Comp and Rel)}
\textbf{Correlation with human Likert score annotations:} We compute the Spearman and Kendall rank correlations between the human-annotated Likert scores and the metric values. 
These values were gathered for the 300 \texttt{CodeReviewer} test instances mentioned in section~\ref{sec:phase1}, (results in Table~\ref{tab:metric_correlations}). 
We exclude human annotations done on the ground truth references (``Ground Truth'' row in Table~\ref{tab:human_study_phase2}) because the reference-based metric value for these would be 1 by default unfairly lowering their correlations.
We observe that while almost all the reference-based metrics have weak to no correlations with human judgment our metric $Rel$ and LaaJ-GPT achieve the greatest correlations.
We also show correlations between human Likert scale annotations for all dimensions and all metrics (including $Con$ and $Comp$) in Table~\ref{tab:all_dim_correlations}.
\\
\textbf{Comparing system rankings:} Arguably the most important desideratum for our metric is the ability to rank systems similar to human evaluators.
We compare the system rankings produced by our metric $Rel$ with rankings produced by human annotations for relevance, showing the correlations in Table~\ref{tab:system_ranking_correlations}.
Only the LaaJ metrics and our metric achieve a strong, statistically significant correlation with the rankings computed from human annotations. 
The system rankings (shown in Table~\ref{tab:system_rankings}) reveal that our metric gets the ranking mostly right, except for LLaMA-3-8B-Instruct, which is ranked slightly lower by our metric compared to human relevance annotations.
We also report the quality dimension scores for each system according to our metrics ($Con$, $Comp$, and $Rel$) and human annotations in Table~\ref{tab:human_study_phase2}. 
Our metrics also have a similar spread of values to the human annotations and greater sensitivity to human preferences as shown in Figure~\ref{fig:metric_monotonicity} and Table~\ref{tab:all_results}.

\subsection{\texttt{CodeReviewer} Dataset Reference Quality}
\label{results:ref_quality}
We compare the quality of the \texttt{CodeReviewer} reference reviews with the reviews generated by the 9 system evaluated by the human annotators. 
The average scores for conciseness, comprehensiveness, and relevance attained by the \texttt{CodeReviewer} references are 3.05, 1.88, and 2.13, while the average scores obtained by all 9 systems are 2.57, 1.84, and 1.99. 
This suggests that the average reference review is barely better than the average evaluated system according to the human annotators for relevance, but they are more concise.
Additionally, the best system according to human annotators (as evident from Table~\ref{tab:human_study_phase2}), GPT-3.5 achieves average scores of 3.63, 2.65, and 2.9 for each dimension -- much better than the reference reviews.
This provides further motivation for the development of reference-free evaluation metrics like \texttt{CRScore} for code review.     

\subsection{Failure Cases}
\label{sec:failure_case_analysis}
We analyze the cases where our metric greatly overestimates or underestimates the quality of a review with respect to human annotations.
We find such cases using the procedure described in Appendix~\ref{sec:failure_case_identification}.
This is also supported by the results in Table~\ref{tab:STS_eval} that show that the STS model used while having a decent amount of recall, has relatively low precision values for most of the thresholds evaluated including $\tau_{best}$, the threshold used in Table~\ref{tab:system_ranking_correlations} and \ref{tab:metric_correlations}.
For underestimation cases, we observe our pseudo-reference generation pipeline generates fewer references on average (2.44) compared to the whole data (4.76). 
This suggests that even though the STS model has a high recall, having fewer pseudo-references, makes it harder to evaluate the relevance of reviews since CRScore might underestimate comprehensiveness compared to humans.
Additionally, we observe reviews like ``Why do we need these imports'' which are brief, contain stopwords, and have fewer relevant tokens, making it hard for STS to recognize their relevance to pseudo-reference.
For overestimation cases, we observe the presence of inline code snippets at a higher rate (45\%) compared to all reviews (28\%) and underestimation cases (12\%).
Some example reviews for each case are shown in Table~\ref{tab:underestimation_cases}, \ref{tab:overestimation_cases}.

\begin{table}[]
\begin{tabular}{@{}lrrr@{}}
\toprule
\textbf{STS Threshold ($\tau$)} & \textbf{P} & \textbf{R} & \textbf{F1} \\ \midrule
0.6 & 0.4055 & 0.9512 & 0.5686 \\
0.65 & 0.4433 & 0.8794 & 0.5895 \\
0.6576 ($\tau_{GT}$) & 0.4475 & 0.8665 & 0.5902 \\
0.7 & 0.4832 & 0.77 & 0.5938 \\
0.7314 ($\tau_{best}$) & 0.5173 & 0.6899 & 0.5913 \\
0.75 & 0.5401 & 0.633 & 0.5828 \\
0.8 & 0.6223 & 0.444 & 0.5183 \\ \bottomrule
\end{tabular}
\caption{Precision (P), Recall (R), and f1-score (F1) achieved by the STS embedding model for various STS thresholds ($\tau$) evaluated on the human annotations linking each review sentence to the pseudo-references addressed by it.}
\label{tab:STS_eval}
\end{table}
\section{Discussion}
In this work we identify issues with current code review evaluation benchmarks like \texttt{CodeReviewer} \cite{CodeReviewer}, which fail to capture the one-to-many nature of code review and contain noisy references.
To enable auditing current evaluation metrics and aid the development of a better metric, we propose three review quality dimensions -- conciseness, comprehensiveness, and relevance based on current literature on reference-free evaluation \cite{mehri-eskenazi-2020-usr, piorkowski2020towards, turzo2024makes}.

To ground these dimensions in topics that reviews should address \cite{rasheed2024ai}, we propose an automated pseudo-reference generation pipeline that leverages LLMs and code smell detectors.
We validate the quality of these pseudo-references via human evaluation.
Based on these dimensions, we develop reliable guidelines for coding review quality using pseudo-references and collect annotations for 9 review generation systems and ``ground truth'' reviews for the \texttt{CodeReviewer} dataset spanning Python, Java, and Javascript.

The collected annotations show that current reference-based metrics except for LLM-as-a-judge with powerful closed source models like GPT-4o fail to capture human preferences, which is further compounded by humans preferring some models over the references.
We propose \texttt{CRScore} as a metric to capture the three dimensions using the pseudo-references generated by open source LLMs and static analysis tools like code smell detectors through STS models like sentence transformers \cite{reimers2019sentence}. 
Our approach has the second greatest alignment with human preferences, lagging only behind LaaJ-GPT and the greatest alignment among open source metrics. 
It achieves the best review quality correlation scores (0.4577 $\tau$ and $0.5425$ $r_s$), system ranking correlations (0.95 $r_s$ and 0.8889 $\tau$), as well as the greatest sensitivity (Figure~\ref{fig:metric_monotonicity}) among open source metrics.
Additionally, it greatly improves over using the Magicoder LLM directly as a judge, while also being more efficient (since the LLM needs to be run only once to produce the pseduo-references as compared to the LLM-as-a-judge, where it needs to be re-run for every review model evaluated). 
However, we also note that despite its reproducibility, efficiency and greater alignment with human preferences, CRScore exhibits only moderate correlations with review quality, and systematically underestimates or overestimates it in some cases. 

\section{Conclusion}
Our work takes the first steps towards addressing the challenges involved in evaluating the quality of code reviews.
We propose useful dimensions for capturing review quality, and offer \texttt{CRScore} as an automated, efficient, open-source and reproducible metric for capturing them. 
We collect a dataset of human judgment of review quality scores to show the validity of \texttt{CRSore}.
We compare \texttt{CRScore} with 7 reference-based metrics, for 9 review generation systems using the collected annotations.
Our metric achieves the best alignment with human preferences among open source metrics and is the most sensitive. 
However, there is scope for improvement by developing better pseudo-reference generation and STS matching methods to try and match the performance of closed-source models like GPT-4o in judging review quality.

\section{Future Work}
Although our metric is a great first step towards reference-free evaluation of code reviews, it still suffers from systematic under and over-estimation errors in certain cases, which causes it to fall short of powerful closed source models.
We believe these limitations stem from pseudo-reference coverage issues and the limitations of STS methods when it comes to matching data containing both code and text.
Future work can extend the pseudo-references by adding components for detecting code security, efficiency issues, etc.  
Also, better embedding models should be developed to measure how faithfully review sentences capture the pseudo-references. 
Additionally, code smell detection can be expanded beyond Python, Java, and Javascript to languages like C/C++, Ruby, and Go, present in the \texttt{CodeReviewer} dataset.

\section*{Limitations}
\begin{itemize}
    \setlength\itemsep{0em}
    \item 
    While annotating reviews for the review quality dimensions of conciseness, comprehensiveness, and relevance, the human annotators are given a list of pseudo-references that overlap with the one used by \texttt{CRScore}.
    We believe this isn't a source of anchoring bias because the human annotators are allowed to add and remove claims from the pseudo-references in the first stage of annotation (while rating pseudo-reference quality).
    The final list of pseudo-references used for review quality annotations is different from the one used by \texttt{CRScore}.
    Additionally, the pseudo-references provide a common ground for the annotators to rate review quality and comprehensiveness. 
    This also helped us achieve high reliability for coding review quality (section~\ref{sec:phase2}).
    \item Semantic textual similarity (STS) models are imperfect at matching relevant pseudo-references to the review sentences, especially when there are very few claims and the reviews contain inline code snippets, where the latter can inflate the STS scores. 
    \item Our pseudo-reference generation pipeline is not comprehensive enough in some cases which can lead to underestimation of review quality as shown by our failure case analysis.
    Additionally, it could be extended by adding more modules like code smell detectors for aspects like code security, code efficiency, etc. similar to \citet{rasheed2024ai}.
    Also, code smell detector tools can be added for languages other than Python, Java, and Javascript, like Go, C/C++, and Ruby present in the \texttt{CodeReviewer} dataset.
    \item Our metric only achieves a moderate correlation with human annotations of review quality, and while it is much better than the reference-based metrics in terms of alignment with human judgment and sensitivity to review quality as judged by humans, it is only a first step towards developing better metrics for code review. Future work should try to address the limitations of our claim generation pipeline and STS methods.
\end{itemize}

\section*{Ethics Statement}
Our work doesn't violate any ethical guidelines and is compliant with copyright rules and regulations as we use an existing publicly available dataset and augment it with annotations of review quality using reviews generated by 9 systems and the references in the dataset.
While there is a slight risk of harmful or toxic text being a part of the pseudo-references generated by the LLM component in our pseudo-reference generation pipeline we don't believe it to be a major risk based on the annotations done for the pseudo-reference quality.


\bibliography{references}
\bibliographystyle{acl_natbib}

\appendix
\label{sec:appendix}
\section{Introduction Details}
This section contains details to corroborate some of the points made in the introduction section.

\subsection{Benefits of Neuro Symbolic Pseudo-Reference Generation}
\label{sec:benefits_of_neurosymbolic_combination}
While LLMs have recently shown promise for evaluating natural language generation (NLG) \cite{li2024leveraging} they suffer from biases like favoring their own generations (``self-selection bias'') \cite{panickssery2024llm} or in other words, if we were to have Magicoder or GPT-3.5 as the evaluator LLM it would assign higher scores to the text generated by Magicoder and GPT-3.5 respectively. 
Code analysis tools (CATs) on the other hand are limited in scope compared to LLMs in detecting issues like best practice violations \cite{AutoCommenter} but don't have any self-selection bias. 
However, combining these methods can reduce the self-selection bias of LLMs, while supplementing the narrow coverage of code analysis tools.
Indeed the results show that despite using Magicoder as the evaluation LLM, our metric \texttt{CRScore} doesn't preferentially rank Magicoder above any models other than LLaMA-3 when compared to the human ranking.

\begin{table}[!tbh]
\centering
\begin{tabular}{ll}
\hline
\textbf{Review} &
  \textbf{\begin{tabular}[c]{@{}l@{}}Missing \\ Context\end{tabular}} \\ \hline
\begin{tabular}[c]{@{}l@{}}Don't redefine, just import the \\ existing one in \textbf{cmdline.py} . :)\end{tabular} &
  \begin{tabular}[c]{@{}l@{}}Folder \\ structure,\\ Codebase \\ organization\end{tabular} \\ \hline
\begin{tabular}[c]{@{}l@{}}I think we can remove this \\ function, right? (duplicate with \\ \textbf{ses\_starter.py})\end{tabular} &
  \begin{tabular}[c]{@{}l@{}}Folder \\ structure,\\ Codebase \\ organization\end{tabular} \\ \hline
\begin{tabular}[c]{@{}l@{}}MPRester(os.environ(\\{[}"MP\_API\_KEY"{]}) can be \\left simply as MPRester() and \\ it will pick up the API key \\from the environment. What \dots \end{tabular} &
  \begin{tabular}[c]{@{}l@{}}Class \\ definition,\\ Environment \\ variables\end{tabular} \\ \hline
\end{tabular}
\caption{Ground truth reviews in the automatically mined \texttt{CodeReviewer} data that assume contextual information about the code base not available in the dataset}
\label{tab:missing_context_eg}
\end{table}


\begin{figure*}[!tbh]
    \centering
    \includegraphics[width=\textwidth]{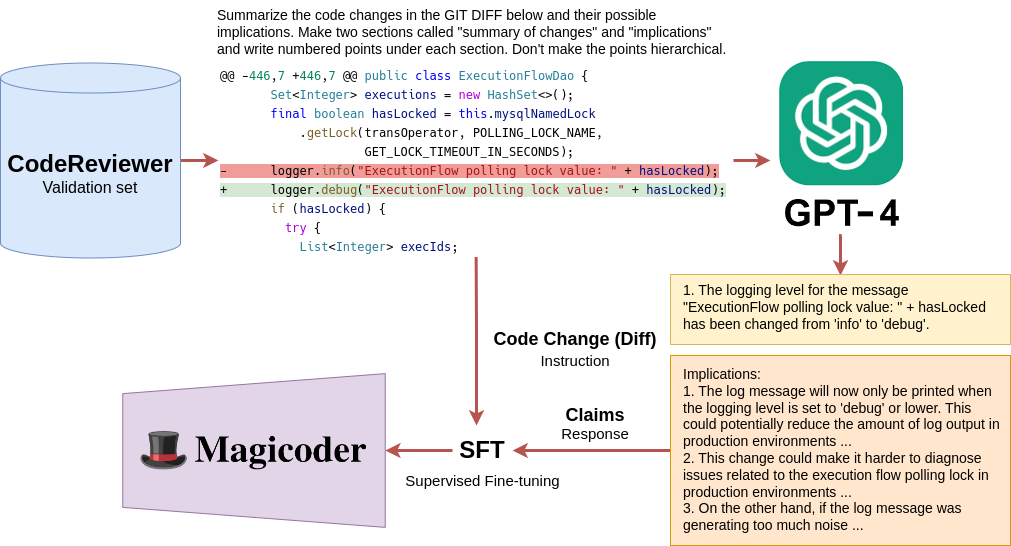}
    \caption{Supervised fine-tuning pipeline for training Magicoder-6.7B for claim generation. We generate synthetic data by using GPT-4 to generate claims for the code changes in \texttt{CodeReviewer} validation set.}
    \label{fig:claim_gen_pipeline}
\end{figure*}

\subsection{Release of Code and Data}
We plan to make the human preferences annotation dataset and code for the evaluation metric and evaluated models public if the paper is accepted. We will add the GitHub link to the code base and the dataset to the camera-ready submission.
\section{More Related Work}
\subsection{Code Specific Reference Based Metrics}
Due to the popularity and convenience of automated reference-based metrics like BLEU \cite{BLEU}, ROUGE \cite{ganesan2018rouge}, and BERTScore \cite{BERTScore} the research community has developed several code-specific versions like CodeBLEU \cite{CodeBLEU}, RUBY \cite{RUBY}, CrystalBLEU \cite{CrystalBLEU} and CodeBERTScore \cite{CodeBERTScore}.
CodeBLEU extends BLEU by incorporating code structure through dataflow and syntax match between generated code and references, while CrystalBLEU filters out trivially shared n-grams.
RUBY incorporates the distance between the syntax tree and program dependency graph of references and generated code. 
CodeBERTScore extends the embedding-based BERTScore by replacing BERT with a pre-trained CodeBERT \cite{CodeBERT} model.
However prior studies have shown metrics like BLEU to have low validity (overlap with human judgment) and reliability for text generation \cite{reiter-2018-structured}, code generation \cite{out_of_the_bleu}, and code migration \cite{RUBY}.
However, metrics like ROUGE, BERTScore, and CodeBERTScore all have a notion of precision, recall, and f-score which is captured by conciseness, comprehensiveness, and relevance respectively.

\subsection{Code Review Automation}
Due to the high time and resource demands of code review automated approaches have gained popularity \cite{yang2024survey}. 
\citet{PORNPRASIT2024107523, LLaMAReviewer, dong2024gpt, fan2024exploring, finetune_code_review_comprehensibility} propose fine-tuning and prompt engineering approaches to leverage LLMs for code review and code-change related tasks.
\citet{Google_Code_Review, rahman2024automating} propose methods for code refactoring based on review comments. 
\citet{AutoCommenter} propose the detection of ``best practice violations'' .
\citet{lin2024improving} propose oversampling reviews from experienced reviewers as a proxy of review quality improving informativeness and correctness of generated reviews.
\citet{rasheed2024ai} propose an LLM agent for code review and code smell detection.

\subsection{Comparing with Review Usefulness/Presentation Style Methods}
\label{sec:additional_related_work:review_presentation_analysis}
Our approach evaluates review informativeness and content related to issues such as code smells compared to prior work which has focused more on linguistic features, style, and presentation. \cite{EvaCRC} proposes EvaCRC an automated approach for review quality evaluation by categorizing reviews along quality attributes like emotion, question, suggestion, and evaluation and rating quality along those attributes on a four-tier grading scale. 
While their conceptualization is similar to ours, they focus on quality attributes related to review style and understandability rather than the actual content. 
Additionally, they don't ground their assessments for these dimensions into a list of claims and issues which makes it harder to understand the scores.
\citet{RevHelper} propose RevHelper as an approach for predicting the usefulness of reviews based on a mixture of three reviewer expertise (e.g. Code Authorship) and five textual features: stop word ratio, reading ease, question ratio, and conceptual similarity. 
Most of the textual features identified by RevHelper deal with review style and presentation, and as pointed out in section~\ref{sec:failure_case_analysis} some of them like code elements could lead to overestimation of review quality. 
While the notion of conceptual similarity comes close to the conceptualization of our $Rel$ score their operationalization of it is very different from ours. 
They directly measure similarity over lines of the code change and review comments.
We believe this is redundant with the notion of preferring reviews that reference code elements and can be gamed by review generation models that reference code snippets incorrectly. 
\citet{CommentBERT} develop CommentBERT an approach for automatically classifying review comments based on their purpose and the type of change required using a taxonomy of comments with 12 categories with dimensions like code\_design, code\_logic, etc.  
While they do classify the review content, they don't look at what makes them useful.

\subsection{Code Smell Detection}
The problem of detecting ``code smells'' or symptoms of design flaws and bad practices (also called \textit{anti-patterns}) has been traditionally tackled by analysis-based approaches \cite{JDeodorant, Paiva2017, liu2017detection}.
Recent work has explored learning-based methods for potentially more nuanced detection of code smells.
\citet{PyCodeSmell} create a dataset of 1k Python code smells like ``Long Method'' and ``Large Class'' to train traditional ML models like random forests. \citet{CodeSmellDLTL} leverage deep learning models like autoencoders and transfer learning for adapting to unseen programming languages.
\citet{liu2024prompt} propose a prompting-based approach for ``Long Method'' and ``Long Parameter List'' code smells in Java.
\citet{rasheed2024ai} create an LLM agent to detect code smells in repositories. 
However, these approaches are limited in the types of smells they target,
training data requirements 
\cite{zhang2024datapreparationdeeplearning},
or lack comprehensive evaluation.
Also, code smells differ across programming languages \cite{code_smells_multi_ling}, and transfer learning approaches can only be leveraged for similar languages \cite{CodeSmellDLTL}.
Due to these limitations of learning-based methods we use traditional language-specific code analysis tools.
\section{Method Details}
This appendix contains additional details on the implementation of our CRScore metric.

\subsection{Distribution of Sentence Similarity Scores}
\begin{figure*}[!tbh]
    \centering
    \includegraphics[width=\textwidth]{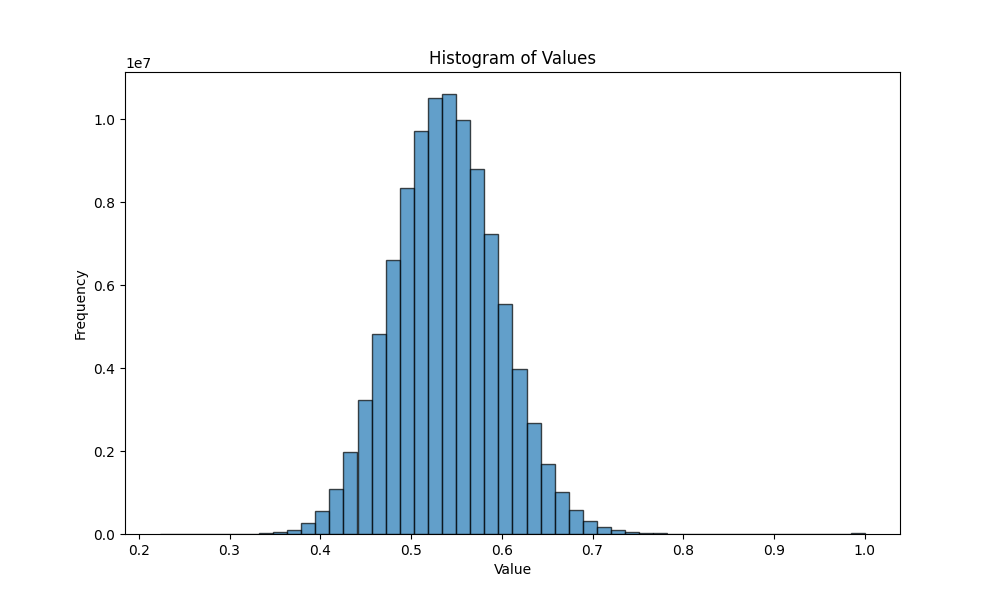}
    \caption{Histogram of sentence similarity of randomly sampled 100K sentence pairs from the \texttt{CodeReviewer} test set showing the scores are roughly normally distributed, justifying the usage of the 5-sigma rule for coming up with the threshold of 0.85 for high similarity used in metric computation.}
    \label{fig:hist_plot_sts}
\end{figure*}

\begin{figure*}[!tbh]
    \centering
    \includegraphics[width=\textwidth]{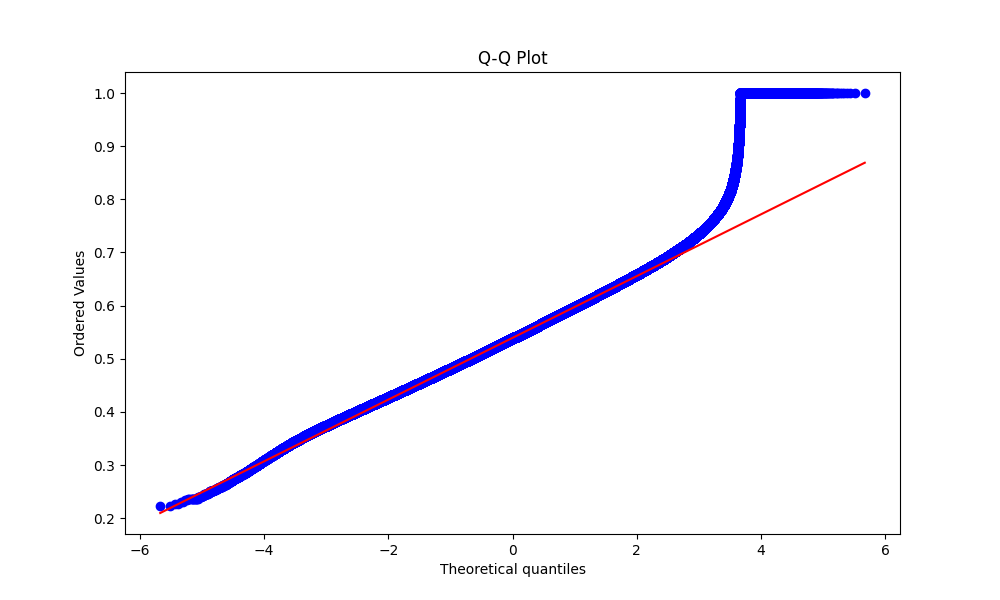}
    \caption{Q-Q plot comparing quantiles of empirically observed sentence similarity scores computed over 100K sentence pairs from the \texttt{CodeReviewer} test set showing the theoretical quantiles match a normal distribution except for really high values. The discrepancy seen here is likely due to the random sample being a smaller subset of the whole 100M+ sentence pairs for which we compute similarities.}
    \label{fig:qq_plot_sts}
\end{figure*}

We plot the histogram of values of the sentence similarity scores in Figure~\ref{fig:hist_plot_sts} showing a roughly normal distribution.
We also plot the quantile-quantile (Q-Q) plot in Figure~\ref{fig:qq_plot_sts} that compares the quantile of a normal distribution with the empirically observed distribution of sentence similarity scores. 
Ideally, the Q-Q plot should be a straight line (shown in red) but we observe deviation towards really high ranges among the actual values (shown in blue). 
Due to computational constraints, these plots are constructed out of a randomly sampled subset of 100k similarity scores from the original 100M+ sentence similarity scores computed from the \texttt{CodeReviewer} test set review pairs.

\begin{figure*}[!tbh]
    \centering
    \includegraphics[width=\textwidth]{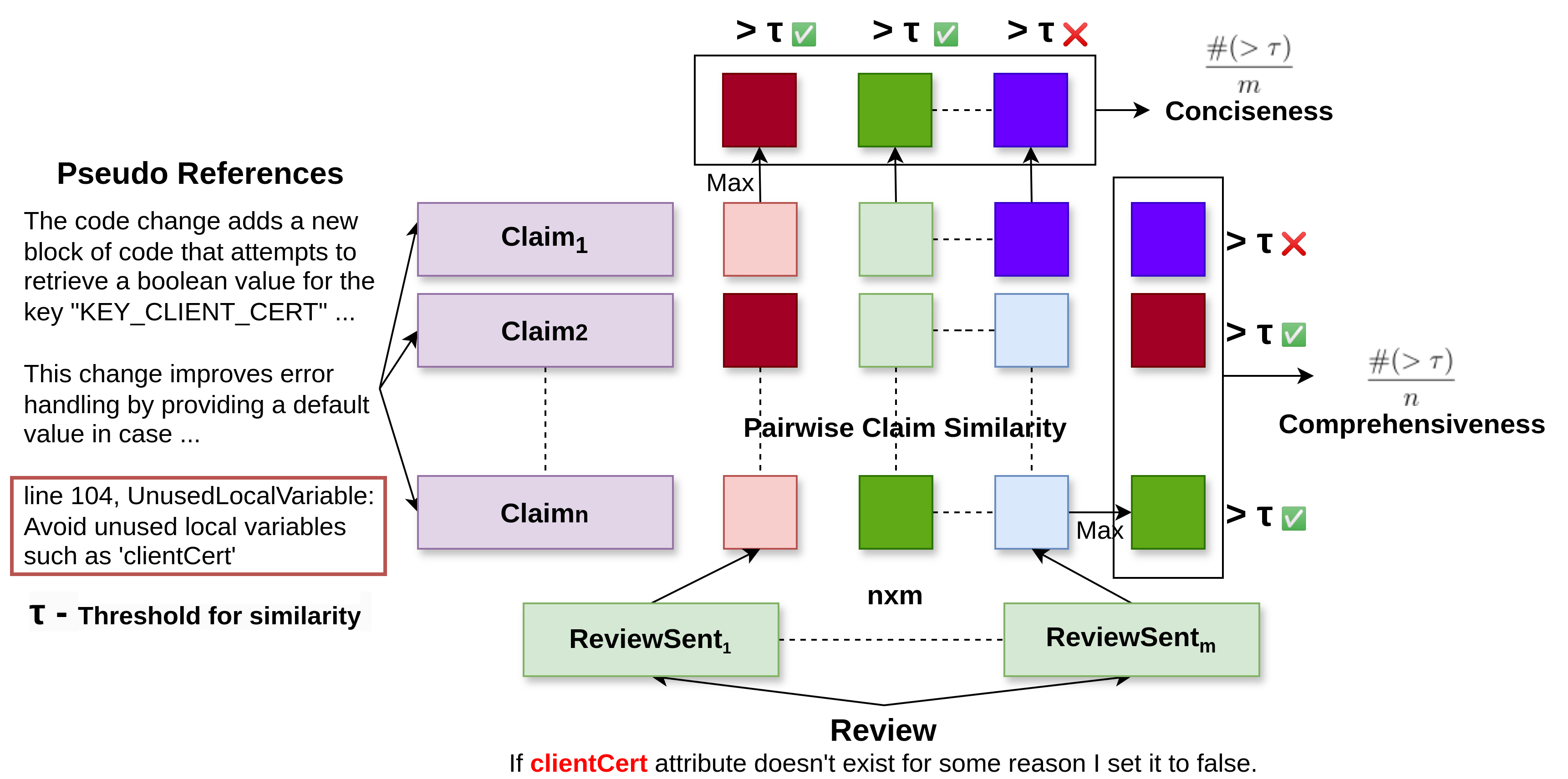}
    \caption{This figure shows how semantic textual similarity (STS) is used to measure the coverage of pseudo-references by the review sentences. We compute pairwise semantic similarities between all pseudo references and review sentences and employ a threshold to compute comprehensiveness as the fraction of pseudo references for which at least one review sentence has higher similarity than the threshold. Meanwhile, conciseness is the fraction of review sentences which high have higher similarity than the threshold with any pseudo reference.}
    \label{fig:metric_computation}
\end{figure*}


\subsection{Code Smell Detection Details}
In this section, we cover some of the details of the code smell detectors used in this study.

\subsubsection{Class Cohesion}
\label{sec:class_cohesion}
Class cohesion captures the degree to which the elements of a class belong together \cite{wikipedia_contributors_2019}. In other words, cohesion measures the strength of the relationship between pieces of functionality (attributes and methods) within a given class. For example, in highly cohesive classes functionality is strongly related and methods and attributes are more co-dependant and hang together as a logical whole \cite{mschwager_2016}.

\subsubsection{Cyclomatic Complexity}
\label{sec:cyclomatic_complexity}
Cyclomatic complexity is a software metric used to indicate the complexity of a program \cite{wikipedia_contributors_2019b}. It is a quantitative measure of the number of linearly independent paths through a program's source code.

Popular tools like Radon\footnote{\url{https://pypi.org/project/radon/}}, a cyclomatic complexity computation tool for Python often resort to a rank-based system that categorizes code blocks based on their cyclomatic complexity and associates them with letter grades as shown in Table~\ref{tab:code_complexity_ranks}. Here ``A'' is the best grade and ``F'' is the worst grade. Code blocks are flagged for code smells if they have a complexity higher than or equal to ``C'' in Table~\ref{tab:code_complexity_ranks}.

\begin{table}[!tbh]
\centering
\resizebox{0.5\textwidth}{!}{%
\begin{tabular}{@{}lll@{}}
\toprule
\textbf{CC Score} & \textbf{Rank} & \textbf{Risk}                           \\ \midrule
1-5               & A             & low - simple block                      \\
6-10              & B             & low - well structured and stable block  \\
11-20             & C             & moderate - slightly complex block       \\
21-30             & D             & more than moderate - more complex block \\
31-40             & E             & high - complex block, alarming          \\
41+               & F             & very high - error-prone, unstable block \\ \bottomrule
\end{tabular}
}
\caption{Code complexity ranks, associated score thresholds, and descriptions of the potential risks.}
\label{tab:code_complexity_ranks}
\end{table}

\subsection{Hyperparameters for Training Pseudo-Reference Generation Pipeline}
We train the \texttt{ise-uiuc/Magicoder-S-DS-6.7B}, with flash attention, random seed of 42, evaluation every 100 steps (eval\_steps), max length padding, train-val split of 0.1 (10\% data used for validation), batch size of 2, maximum training sequence length of 1500 and 5 training epochs.
The training is done on a single A100, 80 GB GPU using the Magicoder training script\footnote{\url{https://github.com/ise-uiuc/magicoder/blob/main/src/magicoder/train.py}}.
\section{Experimental Details}
Further experimental details like guidelines for annotation of pseudo-reference accuracy, review quality, etc.

\subsection{Codebook for Rating Pseudo-Reference Quality}
\label{sec:codebook_for_rating_pseudo_ref_quality}
The raters are shown model-generated pseudo references (claims and implications) about the ``diff'', which either describe the changes or speculate about potential implications.
The claims are statements about the diff related to what changes took place, while the implications cover the effects of the changes or their interpretations like whether they implement a new functionality or even the potential intent of the developers.
Additionally, the raters are also given source files as context, including the versions of the file before and after the code change captured by the diff.
The raters are expected to refer to them if they need more information than just the diff to judge the accuracy of the pseudo-references.

Given these inputs, the raters are supposed to code each pseudo-reference as shown in Table~\ref{tab:coding_pseudo_ref_quality}. 
In further analysis, we excluded the ambiguous claims (code 2) because they were rarely encountered.

\subsection{Rating Review Quality using Pseudo-References}
\label{sec:rating_review_quality}
The raters are given pseudo-references corresponding to the diff but unlike the pseudo-reference quality annotations, they also include issues/smells detected by the static analysis tools.
The issues span formatting issues, bad programming practices, or more abstract patterns known as ``code smells''. 
Code smells are heuristics or code characteristics associated with deeper problems concerning system design. 
It is important to note that they are not bugs, but rather subjective principles that vary across programming languages, developers, teams, etc.
Given the diff and the pseudo-references, the raters evaluate the quality of the review along the three dimensions --- comprehensiveness, conciseness, and relevance.
The raters are again given access to the source files for context and are asked to rate the quality of 10 reviews per code change that are generated by a diverse set of review generation systems and one of them is also the ground truth review from the \texttt{CodeReviewer} dataset.

The rater's task is to first go through the list of pseudo-references and rate their necessity with respect to the code change. This stage is meant to remove any pseudo-references that are unnecessary or redundant with respect to the whole set. 
The second stage of the annotation is to link/associate various pseudo-references to each of the 10 reviews based on which of them are addressed in the review.
Finally, the raters assign a score on a Likert scale of 1 to 5 to each review for each dimension. Some rules of thumb for assigning each score for each dimension are given in Table~\ref{tab:assigning_quality_scores}.

Raters are also given a helpful mental framework to help with the process of linking pseudo-references to the reviews. 
To explain what it means for a review to address a pseudo-reference, we give an analogy to aspect-based product reviews. The list/set of claims is similar to a list/set of product aspect descriptions, while the code review comment is similar to a product review. For example, for the product description and review below: \\
\textbf{Description of a newer version of a Phone}: \\
- The new phone improves battery life by 50\% (battery life) \\
- The new screen has a higher resolution (screen) \\
\textbf{Review}:
The new screen quality is great but the battery runs out quickly.

The review here is talking about both the screen and battery life aspects of the product so you can say it is addressing both aspects (or claims).

Now we can consider an example from the domain of code review: \\
\textbf{Pseudo-references for code change}: \\
- The code change is in the logging of errors in the response handler. (error logging) \\
- The formatting of the error message has been changed. (formatting change of error message) \\
- The previous formatting used '\%s \%' at the end of the error message, which was removed in the updated code. (\% at the end of the error message) \\
- The change in the formatting of the error message will affect the way errors are logged and displayed. (effect of formatting on error logging) \\ 
\textbf{Code review}:
The \%s in the error message is redundant, and the \texttt{indent=4} in \texttt{json.dumps} is unnecessary.

The review talks about the third claim by mentioning the \%s style string at the end of the error logging.


\begin{table*}[!tbh]
\centering
\begin{tabular}{@{}lrl@{}}
\toprule
Dimension                          & Score & Rule of thumb                                                                                                                                                                                                                                                                                                                                         \\ \midrule
\multirow{5}{*}{Conciseness}       & 1     & none of the review is related to the claims+issues.                                                                                                                                                                                                                                                                                                   \\
                                   & 2     & some of the review is related to the claims+issues                                                                                                                                                                                                                                                                                                    \\
                                   & 3     & roughly half of the review is related to the claims+issues                                                                                                                                                                                                                                                                                            \\
                                   & 4     & most of the review is related to the claims+issues                                                                                                                                                                                                                                                                                                    \\
                                   & 5     & basically the whole review is related to the claims+issues                                                                                                                                                                                                                                                                                            \\ \midrule
\multirow{5}{*}{Comprehensiveness} & 1     & whole review doesn’t cover any of the claims+issues                                                                                                                                                                                                                                                                                                   \\
                                   & 2     & review covers at least 1 claim or issue                                                                                                                                                                                                                                                                                                               \\
                                   & 3     & review covers roughly cover half the claims+issues                                                                                                                                                                                                                                                                                                    \\
                                   & 4     & review covers more than half/most of the claims+issues                                                                                                                                                                                                                                                                                                \\
                                   & 5     & review covers practically all the claims+issues                                                                                                                                                                                                                                                                                                       \\ \midrule
Relevance                          &       & \begin{tabular}[c]{@{}l@{}}relevance score must be between conciseness and\\ comprehensiveness scores\end{tabular}                                                                                                                                                                                                                                    \\
                                   &       & \begin{tabular}[c]{@{}l@{}}if one of the two dimensions is "limiting", i.e. has a very\\ low score or poor quality for a given review then the\\ relevance score should be biased/limited by it.\\ \\ E.g. a highly concise review but with very low \\ comprehensiveness should have a relevance score close\\ to the comprehensiveness\end{tabular} \\ \bottomrule
\end{tabular}
\caption{Some rules of thumb for scoring the quality of reviews for each dimension. These rules are meant as guidelines to calibrate multiple annotators and reduce the impact of learning effects in the early stages of annotation.}
\label{tab:assigning_quality_scores}
\end{table*}

\begin{table*}[!tbh]
\centering
\resizebox{\textwidth}{!}{%
\begin{tabular}{@{}llll@{}}
\toprule
{\textbf{Code}} & {\color[HTML]{000000} \textbf{Description}}                                                                                                                                                                                                             & {\color[HTML]{000000} \textbf{Examples}}                                                                                                                                                                                                      & {\color[HTML]{000000} \textbf{Explanation}}                                                                                                                                                                                         \\ \midrule
{\color[HTML]{000000} 1}             & {\color[HTML]{000000} \begin{tabular}[c]{@{}l@{}}Correct: You can find concrete evidence \\ to validate or confirm a claim (either in \\ the diff, the context/source files, or by \\ looking up domain-specific knowledge \\ on the web)\end{tabular}} & {\color[HTML]{000000} \begin{tabular}[c]{@{}l@{}}Any existing calls to the `push` function \\ will need to be updated to include the \\ new `hash` parameter. This could \\ potentially break compatibility with \\ older code.\end{tabular}} & {\color[HTML]{000000} \begin{tabular}[c]{@{}l@{}}This is true since if the parameters are \\ passed by value and there is a \\ parameter after the hash then its value \\ will be accidentally passed as the hash\end{tabular}}     \\
{\color[HTML]{000000} 0}             & {\color[HTML]{000000} \begin{tabular}[c]{@{}l@{}}Incorrect: You can find concrete evidence \\ to contradict a claim (either in the diff, the \\ context/source files, or looking up \\ domain-specific knowledge on the web)\end{tabular}}              & {\color[HTML]{000000} \begin{tabular}[c]{@{}l@{}}The code changes involve the \\ modification of the parameters passed \\ to the ScalarSpaceEncoder function.\end{tabular}}                                                                   & {\color[HTML]{000000} \begin{tabular}[c]{@{}l@{}}False because ScalarSpaceEncoder is \\ a class and not a function.\end{tabular}}                                                                                                   \\
{\color[HTML]{000000} -1}            & {\color[HTML]{000000} \begin{tabular}[c]{@{}l@{}}Unverifiable: You can’t find evidence \\ to confirm or contradict a claim (even \\ after looking at the diff, source files, or \\ looking up domain-specific knowledge \\ on the web)\end{tabular}}    & {\color[HTML]{000000} \begin{tabular}[c]{@{}l@{}}The changes could potentially affect the \\ performance of the code as the order of \\ the arguments does not matter any longer.\end{tabular}}                                               & {\color[HTML]{000000} \begin{tabular}[c]{@{}l@{}}For this claim, the arguments were \\ being passed by keyword so the \\ order didn’t matter for functional \\ correctness but we can’t comment \\ on the performance\end{tabular}} \\
{\color[HTML]{000000} 2}             & {\color[HTML]{000000} \begin{tabular}[c]{@{}l@{}}Ambiguous: The claim is underspecified \\ and has multiple interpretations making \\ it hard to determine what is to be \\ tested/validated\end{tabular}}                                              & {\color[HTML]{000000} \begin{tabular}[c]{@{}l@{}}The function will now only work with \\ valid elements, preventing potential \\ issues down the line.\end{tabular}}                                                                          & {\color[HTML]{000000} \begin{tabular}[c]{@{}l@{}}It is underspecified (unclear) \\ what “valid elements” means.\end{tabular}}                                                                                                       \\ \bottomrule
\end{tabular}
}
\caption{Guidelines for coding accuracy of pseudo-references with example pseudo-references falling within each category and explanations for why they fall in that category.}
\label{tab:coding_pseudo_ref_quality}
\end{table*}

\subsection{Simple Baseline Implementation Details}
\label{sec:simple_baselines}
We create two simple baselines for code review generation a BM-25, kNN style retriever-based approach, and a seq2seq LSTM style model. 
The implementation details of both approaches are described below:
\\
\textbf{BM-25 retriever:}
The BM-25 retriever retrieves a relevant review by matching the closest code change from the \texttt{CodeReviewer} train set to the code change to be reviewed from the test set. 
For an efficient implementation, we create an inverted index from all the code changes in the train set using the Lucene searcher \cite{lucene_searcher} class from Pyserini \footnote{\url{https://pypi.org/project/pyserini/}}.
\\
\textbf{LSTM reviewer:}
We train a single hidden layer encoder-decoder seq2seq LSTM model with Bahdanau attention \cite{bahdanau2014neural} from scratch on the \texttt{CodeReviewer} training data. 
We train it with an Adam optimizer, and negative log-likelihood loss for about 10 epochs, saving the model with the least loss on the \texttt{CodeReviewer} validation set.

\subsection{LLM-as-a-judge Prompting}
\label{sec:appendix:llm_as_a_judge_prompt}
The system prompt is as follows:
\begin{CustomVerbatim}
You are a highly skilled software engineer who has a lot of experience reviewing code changes. Your task is to rate the relevance of any given code change
\end{CustomVerbatim}

The task-specific, review comment evaluation prompt is as follows:
\begin{CustomVerbatim}
TASK PROMPT: You will be asked to rate the relevance of reviews for given Python, Java, or Javascript code changes. A relevant review is one which is both concise and comprehensive. A concise review contains very little text not related to the code change. A comprehensive review contains all the information about a code change that should be covered by a review. A relevant review is comprehensive while being concise.

Now look at the {lang} code change and review below and score the relevance of the review on a scale of 1 to 5

Code Change: {code_change}

Review: {review}

Your score:
\end{CustomVerbatim}

\subsection{Dataset Statistics}
\label{sec:dataset_stats}
We perform all our annotations on the test set of the \texttt{CodeReviewer} dataset, which contains 10169 samples in total and is publicly available with the Apache 2.0 license\footnote{\url{https://huggingface.co/microsoft/codereviewer}}.
The license is permissive and allows us to use the dataset for research purposes.
We randomly sampled 300 samples, with 100 samples across Python, Java, and Javascript -- the languages we have code smell detectors.
We found a few mislabeled code changes for each language that belonged to a different language, which we discarded.
This led to a final dataset of 99, 98, and 96 code changes for Python, Java, and Javascript respectively. 
Also, we end up with 416, 416, and 399 claims that are evaluated as correct, incorrect, or unverifiable.
For the review dataset, we annotate 2.9k reviews (9 systems + ground truth) for each code change from the previous annotation stage.
For all the correlation experiments we exclude the ground truth review annotations around (300), giving us a dataset of roughly 2.6k reviews corresponding to the automated review generation systems.
As mentioned before we excluded the ground truth while computing correlations and system rankings because reference-based metrics by definition would favor (and perfectly rate) the ground truth, unfairly disadvantaging them for computing correlations.

\subsection{Computational Budget}
For running all the systems mentioned in section~\ref{sec:exp_systems} we utilize an 80 GB A100 for the LLM systems, running them on the 10k \texttt{CodeReviewer} test instances.
For running GPT-3.5 (unfortunately no longer publicly available) we utilized around 10k API calls. 
For the synthetic dataset creation for Magicoder for pseudo-reference generation modeling, we called the GPT-4 model (gpt-4-0613) for 1000 \texttt{CodeReviewer} validation instances. 
For the Magicoder model training, we also use an A100 GPU, for 5 epochs (or approximately 8-10 hrs).

\subsection{Annotator Information}
We hired a graduate student and an undergraduate student with experience in writing Java, Python, and Javascript code. The annotation process took several weeks with additional training provided about code review quality, especially about code smells.
The annotation guidelines were also iteratively improved till the high level of reliability reported here was achieved. 
For the compensation, the annotators took up this project for research credits and thus weren't compensated monetarily. 
They are also co-authors of this project.
They were extensively briefed about the research project and how their data would be used as preference data for comparing evaluation metrics and would be publicly released after anonymization.

\subsection{Ethics Review for Data Collection}
The data collection protocol used in this work was approved by the institutional review board (IRB) of the university the authors are affiliated with.

\begin{figure*}[h]
    \centering
    \includegraphics[angle=90, height=0.9\textheight]{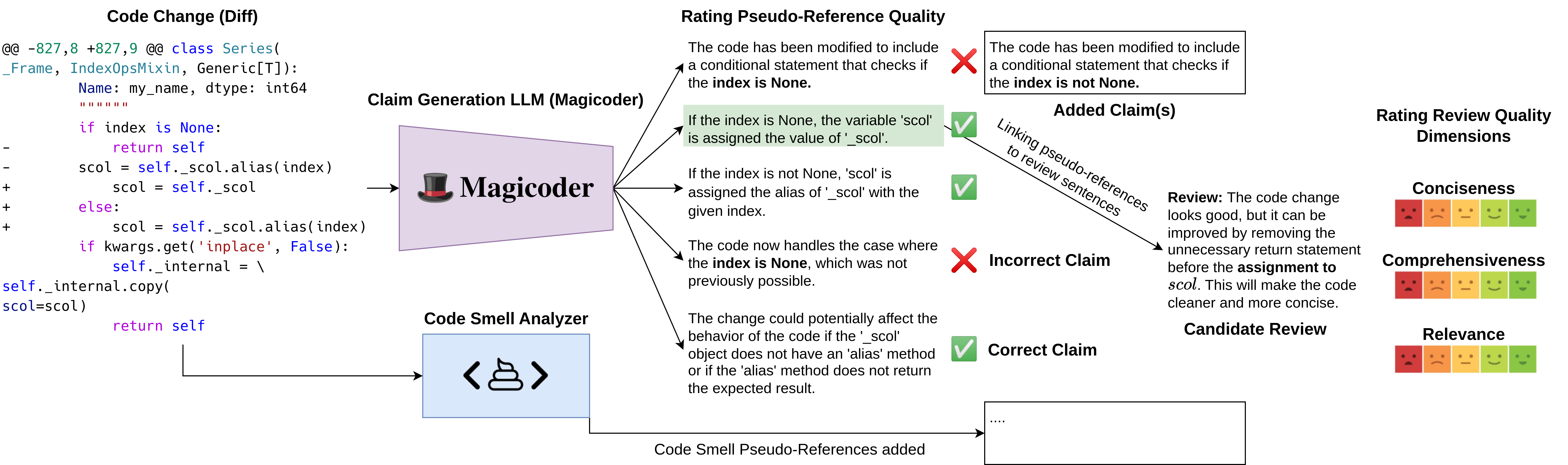}
    \caption{The stages of the annotation pipeline. For the first stage --- Rating Pseudo-Reference Quality the annotators mark correct, incorrect, and unverifiable claims, while also adding any missing claims. In the second stage --- Rating Review Quality Dimensions the annotators are given the updated set of pseudo-references from the first stage along with any pseudo-references generated by the code smell analysis tools to rate candidate reviews on the quality dimensions --- conciseness, comprehensiveness, and relevance on a Likert scale using the pseudo-references.}
    \label{fig:annot_pipeline}
\end{figure*}

\subsection{Does Including the Pseudo-References in the Review Quality Dimension Rating Lead to Biased Annotations?}
\label{sec:appendix:bias_in_phase2}
While at the first glance it might seem that giving the annotators access to the pseudo-reference could bias their operationalization of the review quality dimensions in the favor of CRScore, we believe this is not likely.
Since this study uses a modified set of pseudo-references which is edited by the human annotators to remove incorrect claims and add correct or add missing claims, the set of pseudo-references used by the human annotators is actually different from the automatically generated pseudo-references used by CRScore.
In fact the results demonstrate that the LaaJ-GPT metric which doesn't rely on the pseudo-references can obtain a greater correlation than CRScore with the human review quality dimension annotations despite not having access to the pseudo-references. This demonstrates that the annotations aren't biased towards CRScore. Finally the pseudo-references play a crucial role in helping the annotators concretely assess the dimension of comprehensiveness, since judging the coverage of a review is very hard without a list of things that it should cover. Thus we believe it contributes to the high reliability of the annotations.
\section{Additional Results and Analysis}

\begin{table*}[!tbh]
\centering
\begin{tabular}{@{}lrrrrrr@{}}
\toprule
 & \multicolumn{6}{c}{\textbf{Human Annotations}} \\ \cmidrule(l){2-7} 
\multirow{-2}{*}{\textbf{Metric}} & Con ($\tau$) & Comp ($\tau$) & Rel ($\tau$) & Con ($r_s$) & Comp ($r_s$) & Rel ($r_s$) \\ \midrule
BLEU & 0.0306 & {\color{gray} -0.0227} & {\color{gray} 0.001} & {\color{gray} 0.0358} & {\color{gray} -0.0293} & {\color{gray} -0.0001} \\
\begin{tabular}[c]{@{}l@{}}BLEU\\ (without stopwords)\end{tabular} & 0.0632 & {\color{gray} 0.0221} & 0.0425 & 0.0776 & {\color{gray} 0.0293} & 0.0542 \\
BERTScore & 0.1035 & 0.0622 & 0.081 & 0.1378 & 0.0813 & 1.083 \\
\begin{tabular}[c]{@{}l@{}}Normalized\\ Edit Distance\end{tabular} & {\color{gray} -0.0146} & 0.0443 & {\color{gray} 0.0193} & {\color{gray} -0.0218} & 0.0584 & {\color{gray} 0.0249} \\
\begin{tabular}[c]{@{}l@{}}ROUGE-L\\ F-measure\end{tabular} & 0.0921 & 0.0577 & 0.0757 & 0.1173 & 0.0758 & 0.0989 \\
chrF & 0.1236 & 0.1431 & 0.1484 & 0.1628 & 0.1874 & 0.1966 \\
chrF++ & 0.1294 & 0.1496 & 0.1555 & 0.1707 & 0.1959 & 0.2057 \\
Con ($\tau_{GT}$) ours & 0.4168 &  &  & 0.475 &  &  \\
Comp ($\tau_{GT}$) ours &  & \textbf{0.4982} &  &  & \textbf{0.5832} &  \\
Rel ($\tau_{GT}$) ours &  &  & 0.4437 &  &  & \textbf{0.5405} \\
Con ($\tau_{best}$) ours & \textbf{0.4491} &  &  & \textbf{0.5049} &  &  \\
Comp ($\tau_{best}$) ours &  & 0.4974 &  &  & 0.5754 &  \\
Rel ($\tau_{best}$) ours &  &  & \textbf{0.4567} &  &  & 0.5431 \\ \bottomrule
\end{tabular}
\caption{Kendall and Spearman rank correlations between human annotation for all dimensions: conciseness (Con), comprehensiveness (Comp), and relevance (Rel) and all metrics including our proposed Con, Comp and Rel metrics for both threshold values $\tau_{GT}$ and $\tau_{best}$.}
\label{tab:all_dim_correlations}
\end{table*}

\begin{figure*}[!tbh]
    \centering
    \includegraphics[width=\textwidth]{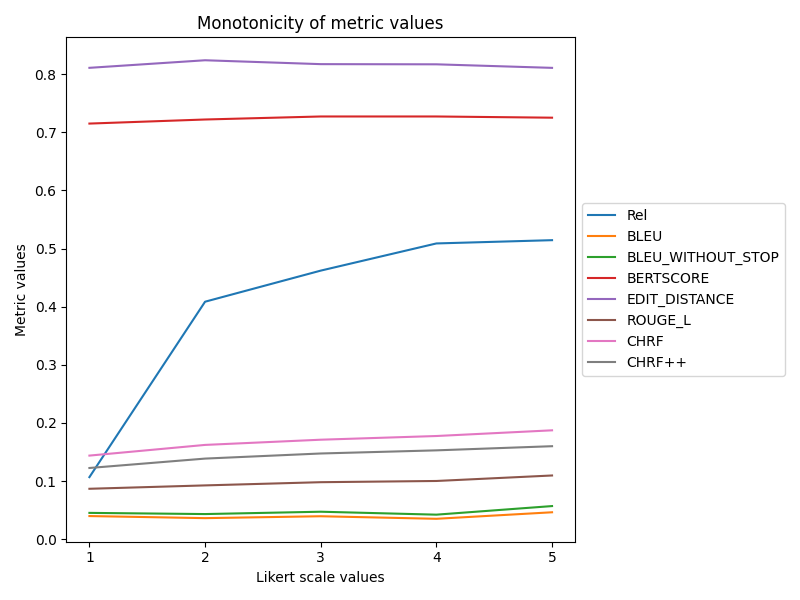}
    \caption{Monotonicity of metric values for reviews of various quality based on the Likert scale human annotations. The Rel metric exhibits the most variation across reviews of different quality while the other metrics have flat plots indicating that they fail to distinguish between reviews of varying quality meaningfully}
    \label{fig:metric_monotonicity}
\end{figure*}

\begin{table*}[!tbh]
\centering
\begin{tabular}{@{}lrrrrrr@{}}
\toprule
\multirow{2}{*}{\textbf{Model}} & \multicolumn{3}{c}{\textbf{Human Annotations}}          & \multicolumn{3}{c}{\textbf{Our Metric}}                 \\ \cmidrule(l){2-7} 
                                & \textbf{Con} & \textbf{Comp} & \textbf{Rel} & \textbf{Con} & \textbf{Comp} & \textbf{Rel} \\ \midrule
BM-25 retriever             & \underline{0.0301} & \underline{0.0112} & \underline{0.0163} &      \underline{0} &      \underline{0} &      \underline{0} \\
LSTM                        & 0.1048 & 0.0361 & 0.0515 & 0.0372 & 0.0123 & 0.0179 \\
CodeReviewer                & 0.5146 & 0.1692 & 0.2311 & 0.4639 & 0.2413 & 0.2974 \\
Stable-Code-3B              & 0.3222 & 0.1383 & 0.1718 & 0.3353 & 0.2091 & 0.2319 \\
DeepSeekCoder-6.7B-Instruct & 0.6108 & 0.3153 & 0.3797 & 0.5037 & 0.3989 & 0.4043 \\
Magicoder-S-DS-6.7B         & 0.4915 & 0.3127 & 0.3351 & 0.4381 & 0.4746 & 0.4032 \\
LLaMA-3-8B-Instruct         & 0.6091 & 0.3368 & 0.4046 & 0.3503 & 0.3967 & 0.3404 \\
CodeLLaMA-13B               & 0.1985 & 0.1546 & 0.1564 & 0.2493 & 0.2528 & 0.2309 \\
GPT-3.5                     & \textbf{0.6564} & \textbf{0.4132} & \textbf{0.4759} & \textbf{0.5622} & \textbf{0.6301} & \textbf{0.5507}  \\
Ground Truth                & 0.5129 & 0.219  & 0.2819 & 0.251  & 0.1598 & 0.1741 \\ \bottomrule
\end{tabular}
\caption{Comparison of our proposed metrics with human annotations for the review quality dimensions. We normalize the human annotated Likert scores per dimension ($DimVal$) from 1 to 5 to 0 to 1 ($NormDimVal$) as: $ NormDimVal = \frac{DimVal -1}{4} $. The results are reported on the subset of 300 annotated \texttt{CodeReviewer} test instances. We highlight the best-performing model (bold) and the worst-performing model (underlined) according to the human annotations and our metric.}
\label{tab:human_study_phase2}
\end{table*}

\begin{table*}[!tbh]
\centering
\begin{tabular}{@{}lrrrrrr@{}}
\toprule
\multirow{2}{*}{\textbf{Metric}} &
  \multicolumn{2}{c}{\textbf{Python}} &
  \multicolumn{2}{c}{\textbf{Java}} &
  \multicolumn{2}{c}{\textbf{Javascript}} \\ \cmidrule(l){2-7} 
 &
  \multicolumn{1}{c}{\textbf{$\tau$}} &
  \multicolumn{1}{c}{\textbf{$r_s$}} &
  \multicolumn{1}{c}{\textbf{$\tau$}} &
  \multicolumn{1}{c}{\textbf{$r_s$}} &
  \multicolumn{1}{c}{\textbf{$\tau$}} &
  \multicolumn{1}{c}{\textbf{$r_s$}} \\ \midrule
BLEU                                                               & \textcolor{gray}{-0.0017} & \textcolor{gray}{-0.0006} & \textcolor{gray}{-0.0101} & \textcolor{gray}{-0.015} & \textcolor{gray}{0.007} & \textcolor{gray}{0.009} \\
\begin{tabular}[c]{@{}l@{}}BLEU \\ (without stopwords)\end{tabular} & 0.0532 & 0.0665 & \textcolor{gray}{0.035} & \textcolor{gray}{0.0442} & \textcolor{gray}{0.0313} & \textcolor{gray}{0.0405} \\
BERTScore & 0.0696 & 0.0942 & 0.0808 & 0.1081 & 0.0846 & 0.1128  \\
\begin{tabular}[c]{@{}l@{}}Normalized\\ Edit Distance\end{tabular} & \textcolor{gray}{0.0367} & \textcolor{gray}{0.046} & \textcolor{gray}{0.0161} & \textcolor{gray}{0.0217} & \textcolor{gray}{0.0109} & \textcolor{gray}{0.0129}  \\
\begin{tabular}[c]{@{}l@{}}ROUGE-L\\ F-measure\end{tabular} & 0.0935 & 0.1201 & 0.0553 & 0.0724 & 0.075 & 0.0981  \\
chrF & 0.1309 & 0.1746 & 0.1573 & 0.2077 & 0.1497 & 0.1982 \\
chrF++ & 0.1387 & 0.1848 & 0.1676 & 0.2208 & 0.1531 & 0.2021 \\
\textbf{Rel ($\tau_{GT}$) (Ours)} & 0.4742 & 0.5788 & 0.3889 & 0.4738 & 0.473 & 0.5738 \\ 
\textbf{Rel ($\tau_{best}$) (Ours)} & 0.4746 & 0.5666 & 0.4093 & 0.4849 & 0.4904 & 0.5816 \\ \bottomrule
\end{tabular}
\caption{Comparing Kendall-Tau ($\tau$) and Spearman Rank ($r_s$) correlation of reference-based evaluation metrics across each language annotated (Python, Java and Javascript) and our reference-free relevance score (Rel (F)) with human annotations for relevance. Correlations that are not statistically significant (p-value < 0.05) are grayed out.}
\label{tab:metric_correlations_by_lang}
\end{table*}

\begin{table*}[!tbh]
\centering
\resizebox{\textwidth}{!}{%
\begin{tabular}{@{}llllr@{}}
\toprule
\textbf{Error Type}                                                 & \textbf{Description}                                                                                                                                  & \textbf{\begin{tabular}[c]{@{}l@{}}Incorrect\\ Reference\end{tabular}}                                                                                                                                                                  & \textbf{\begin{tabular}[c]{@{}l@{}}Corrected\\ Reference\end{tabular}}                                                                                                                                                                                 & \textbf{Frequency (\%)} \\ \midrule
Knowledge Error                                                     & \begin{tabular}[c]{@{}l@{}}Pseudo-reference exhibits\\ incorrect domain knowledge\end{tabular}                                                        & \begin{tabular}[c]{@{}l@{}}The 'optparse' module is being \\ imported with a comment \\ indicating that it is being disabled \\ due to its deprecation.\end{tabular}                                                                    & \begin{tabular}[c]{@{}l@{}}The 'optparse' module is being \\ imported with a comment \\ disabling a pylint \\ deprecated-module warning\end{tabular}                                                                                                   & 6.52                    \\
Reasoning Error                                                     & \begin{tabular}[c]{@{}l@{}}Pseudo-reference exhibits\\ wrong logic applied by the\\ pseudo-reference generator\end{tabular}                           & \begin{tabular}[c]{@{}l@{}}Now, the 'can\_edit\_record' variable \\ is only true if the function \\ 'check\_user\_can\_edit\_record' returns \\ true and the 'format' variable does \\ not start with 't' (in lowercase).\end{tabular}  & \begin{tabular}[c]{@{}l@{}}Now, the 'can\_edit\_record' variable \\ is only true if the function \\ 'check\_user\_can\_edit\_record' returns \\ true and the 'format' variable does \\ not start with 't' (in any lower or \\ uppercase).\end{tabular} & 10.87                   \\
Localization Error                                                  & \begin{tabular}[c]{@{}l@{}}The pseudo-reference generator\\ misunderstands where a\\ code change has taken place\end{tabular}                         & \begin{tabular}[c]{@{}l@{}}The assertion in the test method \\ "test\_idxmapping\_add\_dimension"\\ has been modified.\end{tabular}                                                                                                     & \begin{tabular}[c]{@{}l@{}}The assertion in the test method \\ "test\_idxmapping\_redim" \\ has been modified.\end{tabular}                                                                                                                            & 4.35                    \\
\begin{tabular}[c]{@{}l@{}}Over-generalization\\ Error\end{tabular} & \begin{tabular}[c]{@{}l@{}}The pseudo-reference\\ generator makes an incorrect\\ assumption/generalization from\\ the code change\end{tabular}        & \begin{tabular}[c]{@{}l@{}}The addition of these import \\ statements suggests that the code \\ in this file will now be using the \\ ResLayer and SimplifiedBasicBlock \\ classes from the mmdet.models.utils \\ package.\end{tabular} &                                                                                                                                                                                                                                                        & 19.57                   \\
\begin{tabular}[c]{@{}l@{}}Comprehension\\ Error\end{tabular}       & \begin{tabular}[c]{@{}l@{}}The pseudo-reference seems\\ to "misread" the code change \\ (like thinking removed lines \\ are added, etc.)\end{tabular} & \begin{tabular}[c]{@{}l@{}}The import statement for 'filter', \\ 'range', and 'zip' has been moved \\ from 'scapy.modules.six.moves' to \\ 'scapy.modules.six'.\end{tabular}                                                            & \begin{tabular}[c]{@{}l@{}}'filter' is now also imported \\ from 'scapy.modules.six.moves'\end{tabular}                                                                                                                                                & 58.7                    \\ \bottomrule
\end{tabular}
}
\caption{The various types of errors identified, their descriptions and examples (pseudo-references before and after correction of the error are shown) as well as relative frequencies as percentages are shown here. For this analysis, we annotated 46 erroneous pseudo-references}
\label{tab:pseudo_reference_errors}
\end{table*}

\begin{table*}[!tbh]
\centering
\resizebox{\textwidth}{!}{%
\begin{tabular}{@{}lrrrrrrrrrr@{}}
\toprule
\textbf{Model} & \multicolumn{1}{l}{\textbf{BLEU}} & \multicolumn{1}{c}{\textbf{\begin{tabular}[c]{@{}c@{}}BLEU\\ (without\\ stop)\end{tabular}}} & \multicolumn{1}{l}{\textbf{BERTScore}} & \multicolumn{1}{c}{\textbf{\begin{tabular}[c]{@{}c@{}}Norm.\\ Edit\\ Distance\end{tabular}}} & \multicolumn{1}{c}{\textbf{\begin{tabular}[c]{@{}c@{}}ROUGE\\ L \\ f-score\end{tabular}}} & \multicolumn{1}{l}{\textbf{chrF}} & \multicolumn{1}{l}{\textbf{chrF++}} & \multicolumn{1}{l}{\textbf{\begin{tabular}[c]{@{}c@{}}Con\\ ($\tau_{best}$)\end{tabular}}} & \multicolumn{1}{l}{\textbf{\begin{tabular}[c]{@{}c@{}}Comp\\ ($\tau_{best}$)\end{tabular}}} & \multicolumn{1}{l}{\textbf{\begin{tabular}[c]{@{}c@{}}Rel\\ ($\tau_{best}$)\end{tabular}}} \\ \midrule
BM-25 kNN & 0.036 & 0.043 & 0.718 & 0.805 & 0.069 & 0.134 & 0.113 & 0.002 & 0.001 & 0.001 \\
LSTM & 0.047 & 0.051 & 0.716 & 0.802 & 0.102 & 0.12 & 0.105 & 0.02 & 0.006 & 0.009 \\ \midrule
Transformer\textsuperscript{\textdagger}  & 0.048 &  &  &  &  &  &  &  &  &  \\
T5\textsuperscript{\textdagger}  & 0.044 &  &  &  &  &  &  &  &  &  \\
CodeT5\textsuperscript{\textdagger}  & 0.048 &  &  &  &  &  &  &  &  &  \\
CodeReviewer & 0.054 & \textbf{0.071} & 0.718 & 0.811 & 0.102 & 0.116 & 0.1 & 0.412 & 0.208 & 0.26 \\ \midrule
Stable-Code-Instruct-3B & 0.042 & 0.04 & 0.733 & 0.784 & 0.091 & 0.16 & 0.133 & 0.34 & 0.199 & 0.228 \\
Magicoder-S-DS-6.7B & 0.035 & 0.041 & 0.72 & 0.815 & 0.101 & 0.175 & 0.151 & 0.445 & 0.491 & 0.42 \\
DeepSeekCoder-Instruct-6.7B & 0.045 & 0.054 & \textbf{0.734} & 0.782 & \textbf{0.112} & 0.183 & 0.157 & 0.513 & 0.418 & 0.422 \\ \midrule
CodeLLaMA-Instruct-7B & 0.023 & 0.026 & 0.71 & 0.843 & 0.071 & 0.171 & 0.145 & 0.156 & 0.151 & 0.14 \\
Llama-3-8B-Instruct & 0.014 & 0.016 & 0.699 & \textbf{0.898} & 0.058 & 0.14 & 0.122 & 0.347 & 0.425 & 0.352 \\
LLama-Reviewer\textsuperscript{\textdagger} & \textbf{0.057} &  & \textbf{} & \textbf{} &  & \textbf{} & \textbf{} & \textbf{} & \textbf{} & \textbf{} \\
CodeLLaMA-Instruct-13B & 0.025 & 0.029 & 0.715 & 0.839 & 0.079 & 0.179 & 0.152 & 0.274 & 0.272 & 0.253 \\ \midrule
GPT-3.5-Turbo & 0.037 & 0.044 & \textbf{0.734} & 0.812 & 0.1 & \textbf{0.2} & \textbf{0.171} & \textbf{0.563} & \textbf{0.635} & \textbf{0.558} \\ \bottomrule
\end{tabular}
}
\caption{Results of all eval. metrics and models on the entire test set. All metrics have been normalized to be between 0 and 1. {\textdagger} signifies reported scores. The reference based metrics have a very narrow range of values.}
\label{tab:all_results}
\end{table*}
\begin{table*}[!tbh]
\centering
\resizebox{\textwidth}{!}{%
\begin{tabular}{@{}ll@{}}
\toprule
\textbf{Smell Name}     & \textbf{Description}                                                                                                                                                                                                                          \\ \midrule
Long method             & There exist methods with too many lines (more lines than a set threshold).                                                                                                                                                                    \\
Long parameter list     & There exist methods with more than “n” parameters (n = 6 is used in this study).                                                                                                                                                               \\
Long branch             & When conditional statement branches extend too long or are too nested.                                                                                                                                                                         \\
Many attributes         & When a single class has too many methods or attributes.                                                                                                                                                                                        \\
Many methods            & When a single class has too many methods.                                                                                                                                                                                                     \\
Shotgun surgery         & When a single functionality is fragmented across various classes.                                                                                                                                                                              \\
Class cohesion          & There are some classes with low cohesion (\ref{sec:class_cohesion}).                                                                                                                                                                                                     \\
Code complexity         & \begin{tabular}[c]{@{}l@{}}The code includes blocks with cyclomatic complexity (section \ref{sec:cyclomatic_complexity}) of rank-C \\ or worse (moderate to slightly complex blocks). Please read through the \\ cyclomatic complexity and ranks section for more details.\end{tabular} \\
Long lambda             & \begin{tabular}[c]{@{}l@{}}The code includes lambda functions that exceed a threshold on length \\ (number of characters).\end{tabular}                                                                                                        \\
Long list comprehension & \begin{tabular}[c]{@{}l@{}}The code includes list comprehensions that exceed a threshold on length \\ (number of characters).\end{tabular}                                                                                                     \\ \bottomrule
\end{tabular}
}
\caption{Python code smells detected by the PyScent code smell static analysis tool}
\label{tab:pyscent}
\end{table*}

\begin{table*}[!tbh]
\centering
\begin{tabular}{@{}llll@{}}
\toprule
Rank & \begin{tabular}[c]{@{}l@{}}Human Annotated \\ Relevance\end{tabular} & Rel (ours) & \begin{tabular}[c]{@{}l@{}}chrF++ (best reference \\ based metric)\end{tabular} \\ \midrule
1 & GPT-3.5                     & GPT-3.5                     & GPT-3.5                     \\
2 & LLaMA-3-8B-Instruct         & DeepSeekCoder-6.7B-Instruct & DeepSeekCoder-6.7B-Instruct \\
3 & DeepSeekCoder-6.7B-Instruct & Magicoder-S-DS-6.7B         & CodeLLaMA-13B               \\
4 & Magicoder-S-DS-6.7B         & LLaMA-3-8B-Instruct         & Magicoder-S-DS-6.7B         \\
5 & CodeReviewer                & CodeReviewer                & Stable-Code-3B              \\
6 & Stable-Code-3B              & Stable-Code-3B              & LLaMA-3-8B-Instruct         \\
7 & CodeLLaMA-13B               & CodeLLaMA-13B               & BM-25 retriever             \\
8 & LSTM                        & LSTM                        & CodeReviewer                \\
9 & BM-25 retriever             & BM-25 retriever             & LSTM                        \\ \bottomrule
\end{tabular}
\caption{Rankings of the systems over the 300 human-annotated \texttt{CodeReviewer} instances according to the human-annotated relevance and our relevance metric. The rankings reveal that our method gets the rankings exactly right except for the LLaMA-3-8B-Instruct model (grayed out) which is ranked lower than Magicoder and DeepSeekCoder by our metric but preferred more by humans.}
\label{tab:system_rankings}
\end{table*}

\begin{table*}[!tbh]
\centering
\resizebox{\textwidth}{!}{%
\begin{tabular}{@{}llll@{}}
\toprule
\textbf{Language} & \textbf{Code Change} & \textbf{Claims} & \textbf{Review} \\ \midrule
\multirow{3}{*}{Python} & \multirow{3}{*}{\begin{tabular}[c]{@{}l@{}}\texttt{@@ -24,6 +24,7 @@ CREATE\_TABLE = } \textcolor{olive}{"""}\\ \\  \textcolor{olive}{`complete\_time` datetime DEFAULT NULL,}\\  \textcolor{olive}{`status` enum('SUCCESS','RUNNING',}\\ \textcolor{olive}{'FAILURE',}\\           \textcolor{olive}{'PARTIAL\_SUCCESS','TIMEOUT')}\\ \textcolor{olive}{DEFAULT NULL,}\\ +       \textcolor{olive}{`has\_all\_data` bool DEFAULT NULL,}\\          \textcolor{olive}{`schema\_version` varchar(255)} \\ \textcolor{olive}{DEFAULT NULL,}\\          \textcolor{olive}{`cycle\_timestamp` varchar(255)} \\ \textcolor{olive}{DEFAULT NULL,}\\          \textcolor{olive}{PRIMARY KEY (`id`),}\end{tabular}} & \begin{tabular}[c]{@{}l@{}}A new column named `has\_all\_data` \\ has been added to the table. This \\ column is of boolean data type and \\ allows NULL values by default.\end{tabular} & \multirow{3}{*}{\begin{tabular}[c]{@{}l@{}}\textbf{Magicoder:} This column is \\ not being used anywhere \\ in the codebase. It's a waste \\ of space.\end{tabular}} \\ \cmidrule(lr){3-3}
 &  & \begin{tabular}[c]{@{}l@{}}The addition of the `has\_all\_data` \\ column implies that there might be \\ a need to track whether all the data \\ has been processed or not in the \\ table.This could be used for further \\ data processing or analysis.\end{tabular} &  \\ \cmidrule(lr){3-3}
 &  & \begin{tabular}[c]{@{}l@{}}Existing rows in the table will not \\ have a value for this new column. \\ Therefore, it might require additional \\ data updates or data entry processes \\ to populate this column.\end{tabular} &  \\ \midrule
\multirow{3}{*}{Python} & \multirow{3}{*}{\begin{tabular}[c]{@{}l@{}}\texttt{@@ -14,7 +14,9 @@}\\   \\  \textcolor{purple}{\textbf{\texttt{import}}} \textcolor{teal}{\texttt{abc}}\\  \textcolor{purple}{\textbf{\texttt{import}}} \textcolor{teal}{\texttt{logging}}\\ +\textcolor{purple}{\textbf{\texttt{import}}} \textcolor{teal}{\texttt{datetime}}\\  \textcolor{purple}{\textbf{\texttt{import}}} \textcolor{teal}{\texttt{parameter}}\\ +\textcolor{purple}{\textbf{\texttt{import}}} \textcolor{teal}{\texttt{target}}\\  \textcolor{purple}{\textbf{\texttt{import}}} \textcolor{teal}{\texttt{warnings}}\\  \textcolor{purple}{\textbf{\texttt{import}}} \textcolor{teal}{\texttt{traceback}}\\  \textcolor{purple}{\textbf{\texttt{import}}} \textcolor{teal}{\texttt{pyparsing}} \textcolor{purple}{\textbf{as}} \textcolor{teal}{\texttt{pp}}\end{tabular}} & \begin{tabular}[c]{@{}l@{}}Two new import statements have \\ been added to the code. The first \\ one imports the datetime module, \\ and the second one imports the \\ target module.\end{tabular} & \multirow{3}{*}{\begin{tabular}[c]{@{}l@{}}\textbf{CodeReviewer:} Why do \\ we need these imports?\end{tabular}} \\ \cmidrule(lr){3-3}
 &  & \begin{tabular}[c]{@{}l@{}}The addition of the datetime module \\ suggests that the code may now \\ involve operations related to date \\ and time. This could be for logging \\ purposes, tracking the execution \\ time of the code, or handling \\ dates/times in the program.\end{tabular} &  \\ \cmidrule(lr){3-3}
 &  & \begin{tabular}[c]{@{}l@{}}The addition of the target module \\ indicates that the code may now \\ involve operations related to the \\ target environment or system. \\ This could be for interacting with \\ the target system, or for handling \\ target-specific tasks.\end{tabular} &  \\ \bottomrule
\end{tabular}
}
\caption{Cases where our metric underestimates the relevance of a review by scoring it as zero while the human scores it as 5 (max relevance). We observed that these cases tend to have fewer claims associated with the code change, briefer reviews with few relevant tokens, and fewer inline code snippets.}
\label{tab:underestimation_cases}
\end{table*}

\begin{table*}[!tbh]
\centering
\resizebox{\textwidth}{!}{%
\begin{tabular}{@{}llll@{}}
\toprule
\textbf{Language} & \textbf{Code Change} & \textbf{Claims} & \textbf{Review} \\ \midrule
\multirow{4}{*}{Python} & \multirow{4}{*}{\begin{tabular}[c]{@{}l@{}}\texttt{@@ -971,7 +971,7 @@ \textcolor{blue}{class} \textcolor{teal}{JobTask}\textcolor{blue}{(}\textcolor{teal}{BaseHadoopJobTask}\textcolor{blue}{)}:}\\              \texttt{\textcolor{purple}{\textbf{if}} self.\_\_module\_\_ == \textcolor{olive}{'\_\_main\_\_'}:}\\                  \texttt{\ \ \ \ d = \textcolor{teal}{pickle}.\textcolor{brown}{dumps}\textcolor{blue}{(}self\textcolor{blue}{)}\textcolor{blue}{}}\\\texttt{\ \ \ \ module\_name = \textcolor{teal}{os}.path.\textcolor{brown}{basename}\textcolor{blue}{(}}\\ \texttt{\ \ \ \ \textcolor{teal}{sys}.argv{[}0{]}\textcolor{blue}{)}.\textcolor{brown}{rsplit}\textcolor{blue}{(}\textcolor{olive}{`.'}, 1\textcolor{blue}{)}{[}0{]}}\\ -                \texttt{\ \ \ \ d = d.\textcolor{brown}{replace}\textcolor{blue}{(}b\textcolor{olive}{`(c\_\_main\_\_'}, \textcolor{olive}{``(c''} + module\_name\textcolor{blue}{)}}\\+\texttt{\ \ \ \ d = d.\textcolor{brown}{replace}\textcolor{blue}{(}b\textcolor{olive}{`c\_\_main\_\_'}, b\textcolor{olive}{'c'} + \textbackslash}\\\texttt{\ \ \ \ module\_name.\textcolor{brown}{encode}\textcolor{blue}{(}\textcolor{olive}{`ascii'}\textcolor{blue}{)}\textcolor{blue}{)}}\\ \texttt{\ \ \ \ \textcolor{brown}{open}\textcolor{blue}{(}file\_name, \textcolor{olive}{``wb''}\textcolor{blue}{)}.\textcolor{brown}{write}\textcolor{blue}{(}d\textcolor{blue}{)}}\\  \\              \texttt{\textcolor{purple}{\textbf{else:}}}\end{tabular}} & \begin{tabular}[c]{@{}l@{}}The code change is in the replacement \\ of a part of the byte string 'c\_\_main\_\_' \\ with 'c' followed by the encoded \\ ASCII representation of \\ 'module\_name'.\end{tabular} & \multirow{2}{*}{\begin{tabular}[c]{@{}l@{}}\textbf{CodeReviewer:} Shouldn't this be \\ `module\_name.encode('ascii')`?\end{tabular}} \\ \cmidrule(lr){3-3}
 &  & \begin{tabular}[c]{@{}l@{}}The original code was replacing \\ '(c\_\_main\_\_' with '(c' followed by \\ 'module\_name'.\end{tabular} &  \\ \cmidrule(l){3-4} 
 &  & \begin{tabular}[c]{@{}l@{}}The new code change corrects this \\ by encoding 'module\_name' to \\ ASCII before replacing 'c\_\_main\_\_' \\ with 'c' followed by the encoded \\ 'module\_name'.\end{tabular} & \multirow{2}{*}{\begin{tabular}[c]{@{}l@{}}\textbf{StableCode:} This seems \\ unnecessary, the pickle \\ module already encodes \\ the module name.\end{tabular}} \\ \cmidrule(lr){3-3}
 &  & \begin{tabular}[c]{@{}l@{}}This change could potentially affect \\ the behavior of the code if \\ 'module\_name' contains non-ASCII \\ characters.\end{tabular} &  \\ \bottomrule
\end{tabular}
}
\caption{Cases where our metric overestimates the relevance of a review by assigning it the maximum score while the human scores it at the lowest Likert score value of 1. We observed that these cases tend to have far more inline-code snippets than the rest of the data and underestimation cases.}
\label{tab:overestimation_cases}
\end{table*}

\subsection{Finding Failure Cases of \texttt{CRScore}}
\label{sec:failure_case_identification}
We divide the scores spanned by our metric into 5 equally sized bins (Q1, Q2, Q3, and Q4 being the quantiles) and then find reviews where the metric underestimates the Rel value (value less than Q1 but a human rating of 5) and overestimates the Rel value (value greater than Q4 but a human rating of 1).
We find 16 (0.61\%) and 98 (3.74\%) cases respectively for underestimation and overestimation, moreover, without these cases, our metric ($\tau_{best}$) attains correlations of $\tau=0.5462$ and $r_s=0.6431$ and $\tau=-0.5403$ and $r_s-0.6131=$ for these cases, indicating their influence despite being less than $4\%$ of the data. 

\subsection{Impact of Varying STS Threshold ($\tau$)}

\begin{table}[!tbh]
\centering
\begin{tabular}{@{}lrr@{}}
\toprule
\textbf{STS Threshold ($\tau$)}                                                    & \multicolumn{1}{l}{$\tau$} & \multicolumn{1}{l}{$r_s$} \\ \midrule
$\tau$=0.6 & 0.3975 & 0.4874 \\
$\tau$=0.65 & 0.4572 & 0.5512 \\
$\tau_{GT}$=0.6576 & 0.4437 & 0.5405 \\
$\tau$=0.7 & 0.4433 & 0.5407 \\
$\tau_{best}$=0.7314 & 0.4567 & 0.5431 \\
$\tau$=0.75 & 0.4473 & 0.5277 \\
$\tau$=0.8 & 0.3946 & 0.4539 \\
\bottomrule
\end{tabular}
\caption{Effect of varying STS threshold ($tau$) on the Kendal-Tau and Spearman Rank correlation of CRScore relevance measure (Rel) with human-annotated relevance. We note that our approach is robust to slight variations in the threshold making it robust.
}
\label{tab:STS_threshold_variation}
\caption{}
\end{table}

\end{document}